            \newif\ifdraft \newif\ifcolorfigs \newif\ifJFM  %
            \drafttrue \colorfigstrue \JFMtrue              %
  \draftfalse     

\documentclass{jfm}

\ifJFM
  \usepackage{upmath}
\fi

\usepackage{ifpdf}
\ifpdf
  \usepackage[pdftex]{color,graphicx}
  \usepackage{epstopdf}
\else
  \usepackage[dvips]{color,graphicx}
\fi

\usepackage{float}
\usepackage{natbib}
\usepackage{amsmath,amsfonts,amssymb,amsbsy,amscd,amsgen}
\usepackage{url}
\usepackage{subfigure,overpic}
\usepackage{ifthen}
\usepackage{appendix}
\graphicspath{{../figs/}{../Fig/}}  

\usepackage[active]{srcltx}

\ifcolorfigs
  \usepackage[pdftex,colorlinks]{hyperref} 
\else
\fi

 \newcommand{\NRPOsTot}{48} 

\ifdraft    
   
   \newcommand{\PC}[1]{$\footnotemark\footnotetext{PC: #1}$}
   
   \newcommand{\NBB}[1]{$\footnotemark\footnotetext{NBB: #1}$}
   
   \newcommand{\APW}[1]{$\footnotemark\footnotetext{APW: #1}$}
   
   \newcommand{\KYS}[1]{$\footnotemark\footnotetext{KYS: #1}$}

   \newcommand{\file}[1]{$\footnotemark\footnotetext{{\bf file} #1}$}
   \newcommand{\mycomment}[2]{\noindent \textbf{\underline{#1}}: \emph{#2}}
   \newcommand{\edit}[1]{{\color{blue}#1}} 
\else   

   \newcommand{\PC}[1]{}
   \newcommand{\JFG}[1]{}
   \newcommand{\NBB}[1]{}
   \newcommand{\APW}[1]{}

   \newcommand{\KYS}[1]{}
   
   \newcommand{\file}[1]{}
   \newcommand{\mycomment}[2]{}
   \newcommand{\edit}[1]{#1}               

\fi 

\ifcolorfigs            

\else

\fi

\newcommand{\rf}     [1] {~\cite{#1}}
\newcommand{\rfp}[1] {~\citep{#1}}
\newcommand{\refref} [1] {\cite{#1}}

\newcommand{\refeq}  [1] {(\ref{#1})}

\newcommand{\reffig} [1] {figure~\ref{#1}}
\newcommand{\reffigs} [2] {figures~\ref{#1} and~\ref{#2}}
\newcommand{\refFig} [1] {Figure~\ref{#1}}

\newcommand{\reftab} [1] {table~\ref{#1}}

\newcommand{\refsect}[1] {\S\,\ref{#1}}

\newcommand{\refappe}[1] {appendix~\ref{#1}}

\newcommand{\beq}{\begin{equation}}
\newcommand{\continue}{\nonumber \\ }

\newcommand{\eeq}{\end{equation}}
\newcommand{\ee}[1] {\label{#1} \end{equation}}

\newcommand{\bea}{\begin{eqnarray}}
\newcommand{\eea}{\end{eqnarray}}
\newcommand{\barr}{\begin{array}}
\newcommand{\earr}{\end{array}}

\newcommand{\ie}{{i.e.}}        

\ifJFM                          

\else
\fi

\newcommand{\gSpace}{\ensuremath{{ \phi}}}   

\newcommand\PoincSec{Poincar\'e section}

\newcommand{\po}{periodic orbit}

\newcommand{\rpo}{rela\-ti\-ve periodic orbit}
\newcommand{\Rpo}{Rela\-ti\-ve periodic orbit}

\newcommand{\mslices}{method of slices}

\newcommand{\fFslice}{first Fourier mode slice}
\newcommand{\FFslice}{First Fourier mode slice}

\newcommand{\zeit}{\ensuremath{t}}  
\newcommand{\sliceTan}[1]{\ensuremath{\mathbf{t}{}'_{#1}}}    
\newcommand{\groupTan}{\ensuremath{\mathbf{t}}}    
\newcommand{\LieEl}{\ensuremath{g}}  
\newcommand{\LieElz}[1]{\ensuremath{\LieEl_z (#1)}}  
\newcommand{\id}{{\ \hbox{{\rm 1}\kern-.6em\hbox{\rm 1}}}}

\newcommand{\On}[1]{\ensuremath{\textrm{O}(#1)}}
\newcommand{\SOn}[1]{\ensuremath{\textrm{SO}(#1)}}         
\newcommand{\sspRed}{\ensuremath{{\hat{\ssp}}}}    
\newcommand{\velRed}{\ensuremath{\hat{\vel}}}    
\newcommand{\slicep}{{\ensuremath{\sspRed'}}}   
\newcommand{\Group}{\ensuremath{G}}         

\newcommand{\NS}{Navier--Stokes}
\newcommand{\NSe}{Navier--Stokes equations}

\newcommand{\KS}{Kuramoto--Sivashinsky}
\newcommand{\Reynolds}{\textit{Re}}  

\newcommand{\eqv}{equilib\-rium}

\newcommand{\eqva}{equilib\-ria}

\newcommand{\reqv}{travelling wave}

\newcommand{\reqva}{travelling waves}
\newcommand{\Reqva}{Travelling waves}

\newcommand{\cohStr}{invariant solution}
\newcommand{\CohStr}{Invariant solution}
\newcommand{\ecs}{exact coherent structure}

\newcommand{\stateDsp}{state-space}
\newcommand{\StateDsp}{State-space}
\newcommand{\Statesp}{State space}
\newcommand{\statesp}{state space}

\newcommand{\REQV}[2]{\ensuremath{\mathrm{TW}_{#1#2}}} 

\ifdraft
	\newcommand{\RPORUN}[2]{\ensuremath{\textrm{RPO}_{#1/#2}}}
\else
	\newcommand{\RPORUN}[2]{\ensuremath{\textrm{RPO}_{#1}}}
\fi

\newcommand\flow[2]{{f^{#1}(#2)}}
\newcommand\timeflow{{f^t}}

\newcommand{\PoincS}{{\cal P}}     

\newcommand{\dmn}{-dimensional} 

\newcommand{\expct}    [1]{\left\langle {#1} \right\rangle}

\renewcommand\Im{{\rm Im\,}}
\renewcommand\Re{{\rm Re\,}}

\newcommand{\shift}{\ensuremath{\ell}}
\newcommand\period[1]{{T_{#1}}}         
\newcommand{\pS}{{\cal M}}          
\newcommand{\ssp}{a}            
\newcommand{\vel}{\ensuremath{v}}   

\newcommand{\eigExp}[1][]{
\ifthenelse{\equal{#1}{}}{\ensuremath{\lambda}}{\ensuremath{\lambda^{(#1)}}}
                        }
\newcommand{\eigRe}[1][]{
\ifthenelse{\equal{#1}{}}{\ensuremath{\mu}}{\ensuremath{\mu^{(#1)}}}
                        }
\newcommand{\eigIm}[1][]{
  \ifthenelse{\equal{#1}{}}{\ensuremath{\omega}}{\ensuremath{\omega^{(#1)}}}
            }

\newcommand{\bnabla}{\mbox{\boldmath $\nabla$}}
\renewcommand{\vec}[1]{\mbox{\boldmath $#1$}}

\newcommand{\inprod}[2]{\langle{#1}\vphantom{#2}|\vphantom{#1}{#2}\rangle} 
\newcommand{\inprodE}[2]{\langle{#1}\vphantom{#2}|\vphantom{#1}{#2}\rangle_{L^2}} 
\newcommand{\normE}[1]{\left\| #1 \right\|_{L^2}}
\newcommand{\lowpass}{low pass}
\newcommand{\lowPass}{Low pass}
\newcommand{\inprodLP}[2]{\langle{#1}\vphantom{#2}|\vphantom{#1}{#2}\rangle_{LP}} 
\newcommand{\FloquetV}{\ensuremath{V}}
\newcommand{\FloquetVRed}{\ensuremath{\hat{V}}}

\newcommand{\TWRUN}[2]{\ensuremath{\textrm{TW}_{#1}}}

\title[Relative periodic orbits in turbulent pipe flow]
{
Relative periodic orbits form the backbone of turbulent pipe flow
}
\author[
N.\,B.\,Budanur et al.
        ]
{
N.\ns B.\ns B\ls U\ls D\ls A\ls N\ls U\ls R$^{1,5}$,
\ns
K.\ns Y.\ns S\ls H\ls O\ls R\ls T$^{2}$,
\ns
M.\ns F\ls A\ls R\ls A\ls Z\ls M\ls A\ls N\ls D$^{3,5}$,
\ns
A.\ns P.\ns W\ls I\ls L\ls L\ls I\ls S$^{4,5}$,
\ns
\and
P.\ns C\ls V\ls I\ls T\ls A\ls N\ls O\ls V\ls I\ls \'C$^{2,5}$
}
\affiliation{
$^1$ Institute of Science and Technology (IST), 
Am Campus 1, 3400 Klosterneuburg, Austria
\\[\affilskip]
$^2$School of Physics,
 Georgia Institute of Technology,
 Atlanta, GA  30332, USA
\\[\affilskip]
$^3$Department of Mechanical Engineering,
Massachusetts Institute of Technology,
77 Massachusetts Ave., Cambridge, MA 02139, United States
\\[\affilskip]
$^4$School of Mathematics and Statistics,
University of Sheffield, S3\,7RH, UK
\\[\affilskip]
$^5$ Kavli Institute for Theoretical Physics,
UC Santa Barbara, Santa Barbara, CA 93106
}

\pubyear{2017}
\begin{document}
\maketitle

\ifdraft
    \tableofcontents
\fi

    \abstract{
Chaotic dynamics of low-dimensional systems, such as Lorenz or R\"ossler
flows, is guided by the infinity of periodic orbits embedded in their strange
attractors. Whether this also be the case for the infinite-dimensional
dynamics of Navier--Stokes equations has long been speculated, and is a topic
of ongoing study. Periodic and relative periodic solutions have been shown to
be involved in transitions to turbulence.  Their relevance to turbulent
dynamics---specifically, whether periodic orbits play the same role in
high-dimensional nonlinear systems like the Navier--Stokes equations as they
do in lower-dimensional systems---is the focus of the present investigation.
We perform here a detailed study of {pipe flow} relative periodic orbits with
energies and mean dissipations close to turbulent values. We outline several
approaches to reduction of the translational symmetry of the system. We study
pipe flow in a minimal computational cell \edit{at $Re=2500$}, and report a library of invariant
solutions found with the aid of the method of slices. Detailed study of the
unstable manifolds of a sample of these solutions is consistent with the
picture that relative periodic orbits are embedded in the chaotic saddle and
that they guide the turbulent dynamics.
    }

\section{Introduction}
\label{s:intro}

Revealing the underlying mechanisms of fluid turbulence is a
multidisciplinary endeavour that brings together pure and applied
mathematics, high performance computation, and experimental physics. Over
the past two decades, this effort has led to significant progress in our
understanding of transitionally turbulent fluid flows in physically
motivated geometries, such as a circular pipe. Today we have numerical
evidence that the laminar state of the pipe flow
\citep{Hagen1839,Poiseuille1844} is linearly stable for all cases that
can be observed in laboratory experiments, \ie, for Reynolds numbers up
to $\Reynolds= 10^7$ \citep{MesTre03}.
In addition, both numerical and laboratory experiments
\citep{AVWIHO10,hof2006flt} indicate that turbulence \edit{of finite
spatial extent, either in the form of a localised patch or turbulence or
within a geometry of finite volume,} has a finite lifetime at
transitional \Reynolds\ values.
\edit{(When the spatial expansion of turbulence in larger domains
defeats relaminarisations, such that it persists indefinitely, the system
becomes a strange attractor; \cite{AMdABH11}.)}
These observations suggest, from the
dynamical systems point of view, that the study of a turbulent pipe flow
is the study of a chaotic saddle, \ie, a strange repeller in the
infinite-dimensional \statesp\ of the solutions to \NSe.
For low-dimensional
dynamics, it is known that strange sets are shaped by the `\cohStr s'
and their stable and unstable manifolds.
    \footnote{
    Here by `\cohStr s' or `\ecs s' we mean compact, time-invariant
    solutions that are set-wise invariant under the time evolution and
    the continuous symmetries of the
    dynamics. Invariant solutions include, for instance, \eqva, \reqva, \po s and invariant tori.
    Note in particular that the closure of a \rpo\ is an invariant torus.
    }

This intuition motivated several groups
\citep{FE03,WK04,Pringle07,ACHKW11}
to investigate \cohStr s of \NSe\ in a circular pipe; these studies, in
turn, were followed by experimental observations \citep{science04,DeSo14}
of relatively close visits of the turbulent trajectories to some of the
numerical \reqv\ solutions. With a fast-growing catalogue of exact
{\cohStr s} of the \NSe\ in hand, acquired by our group and others
\citep{WiShCv15,channelflow}, we are nearing the point where focus turns
from finding \cohStr s to constructing their stable and unstable
manifolds, the building blocks of a chaotic saddle.

Most of the early studies of \cohStr s in pipe flow had focused on
structures that play role in transition to turbulence. Typically these
solutions emerged in saddle-node bifurcations (or further bifurcations of
such solutions) as lower/upper-branch pairs. Lower-branch
solutions appeared to belong to \statesp\ regions that separated initial
conditions into those that uneventfully relaminarize, and those that
develop into turbulence. Moreover, these solutions are characterized by
structures smoother than those observed in turbulence, hence their
numerical study was relatively simple and required moderate numbers of
computational degrees of freedom. Upper-branch solutions, on the other
hand, undergo very complex sequences of bifurcations
\rfp{mellibovsky12} upon increasing \Reynolds , giving rise to
complicated dynamics with many of the resulting solutions distant from
the turbulent regions. While these bifurcations are precursors of
turbulence in pipe flow, a complete continuation from upper-branch
solutions to turbulence is a very hard task: Many solutions undergo
sequences of bifurcations in different regions of the \statesp, then
sometimes merge through boundary crises that are hard to detect.

By contrast, the strategy of the present study is to extract \cohStr s from
close recurrences of turbulent flow simulations \citep{pchaot}, for a
given \Reynolds\ and domain geometry, with the aim of identifying
dynamically relevant structures, without any prior knowledge of the
bifurcation sequences from which they might have originated.

Pipe flow is driven by a pressure gradient; hence, all of its
finite-amplitude solutions drift downstream.
The simplest 
\cohStr s in such translationally-invariant systems are \reqva. Due to
the azimuthal-rotation invariance of the pipe flow, in general one
anticipates finding \reqva\ that simultaneously drift downstream and
rotate about the axis of the pipe (rotational waves). Since the motions
of such solutions can be eliminated by a change to the co-moving frame,
moving along the system's symmetry directions, the physical observables
associated with them, such as wall friction or dissipation, do not change
in time. In other words, the dynamical information contained in these
solutions is rather limited. The simplest time-dependent \cohStr s that
capture dynamics in terms of time-dependent, but symmetry-invariant,
observables are the \rpo s, which are velocity field profiles that
exactly recur at a streamwise (downstream) \edit{shifted} location after a finite
time. \edit{More generally, \rpo s may have azimuthal rotations in
			addition to the streamwise drifts, however, such orbits are not
			contained in the symmetry-subspace we study here.}

\edit{
In this work, we present the \NRPOsTot\ \rpo s and $10$ \reqv s,
adding $19$ new solutions to the $29$ solutions reported in~\refref{WiShCv15}.
Six of these new \rpo s are computed by the method of multi-point
shooting (for the first time in the pipe flow context) whereby the
initial guesses for longer orbits are constructed from known shorter
orbits that shadow them (see \refsect{s:multi}).

Next, we investigate the role the invariant solutions play in shaping the turbulent dynamics.
To this end we carry out global and local  \statesp\ visualizations,
both in the symmetry-reduced \statesp, and in its \PoincSec s.
For global visualizations, we take a data-driven approach and
project \rpo s and turbulent dynamics onto `principal components'
obtained from the symmetry-reduced turbulence data.
We show that this approach has only limited descriptive power
for explaining the organization of solutions in the \statesp.
We then move onto examining the unstable manifold of our shortest
\rpo\ and illustrate how it shapes the nearby solutions.
This computation extends \rf{BudHof17}'s method for studying
the unstable manifolds of `edge state' \rpo s to the solutions
that are embedded in turbulence, with unstable manifold dimensions
greater than one. Finally, we demonstrate
that when a turbulent trajectory visits the neighbourhood of this
\rpo, it shadows it for a finite time interval.

Our results demonstrate the necessity of symmetry reduction for
\statesp\ analysis. We reduce the continuous translational symmetry along the pipe by bringing all states to a
symmetry-reduced \statesp\ (the slice), and contrast this with the `method of connections'.
The remaining discrete azimuthal symmetry is reduced by defining a \emph{fundamental domain}
within the slice, where each state has a unique representation.
We demonstrate, on concrete examples, that this symmetry reduction makes
possible a dynamical analysis of the pipe flow's \statesp.

The paper is organized as follows.
In \refsect{s:review} we describe the pipe flow and its symmetries.
In \refsect{s:symmRed} we discuss the {method of slices} used to reduce
the continuous symmetry.
The computed invariant solutions are listed and discussed in
\refsect{s:data}.
In \refsect{s:stateSp} we investigate the dynamical role of the invariant
solutions using global and local \statesp\ visualizations.
Section \refsect{s:concl} contains our concluding remarks.
}

\section{Pipe flow}
\label{s:review}

The flow of an incompressible viscous fluid through a pipe of
circular cross-section is considered.  Fluid in a long pipe carries
large momentum, which in turn smooths out fluctuations in the mass flux
on short time-scales.  We therefore consider flow with constant mass
flux whose governing equations read
\beq
\frac{\partial\vec{u}}{\partial t} + \vec{U}\cdot\bnabla\vec{u} +
\vec{u}\cdot\bnabla \vec{U} + \vec{u}\cdot\bnabla\vec{u} = - {\bnabla} p
+ 32\frac{\beta}{\Reynolds}\hat{{\bf z}} +
\frac{1}{\Reynolds}{\bnabla}^2 \vec{u}\,, \qquad \bnabla \cdot \vec{u} =
0 \, .
\ee{NavStokesDev}
The equations are formulated in cylindrical-\-polar coordinates $(r,
\theta, z)$ denoting the radial coordinate $r$, the azimuthal angle
$\theta$ and the stream-wise (or axial) coordinate $z$ along the pipe.
The Reynolds number is defined as $\Reynolds = U D / \nu$, where $U$ is
the mean velocity of the flow, $D$ is the pipe diameter, and $\nu$ is the
kinematic viscosity. The governing equation~\refeq{NavStokesDev} is
non-dimensionalized by scaling the lengths by $D$, the velocities by $U$,
and time by $D/U$. The velocity $\vec{u}\edit{=(u,v,w)}$ denotes the deviation from the
dimensionless laminar Hagen--Poiseuille flow \eqv\ $\vec{U}(r)=
2\,(1-(2r)^2)\,\hat{\vec{z}}$.
\edit{In addition to the pressure gradient required to maintain
laminar flow, the excess pressure required to maintain
constant mass flux is measured by the feedback variable
$\beta=\beta(\vec{u})$ --- the total dimensionless
pressure gradient is $(1+\beta)(32/Re)$ and $\beta=0$ for laminar flow.}
\edit{The Reynolds number used throughout this work is $Re = 2500$.}

Our computational cell is in the $m_0=4$ rotational subspace,
$\Omega: (r,\theta,z)\in
[0,\frac{1}{2}]\times[0,\frac{\pi}{2}]\times[0,\frac{\pi}{\alpha}]$ with $\alpha=1.7$,
or in wall units for the wall-normal, spanwise and streamwise dimensions
respectively, $\Omega^+\approx[100,160,370]$.
The variables in \refeq{NavStokesDev} are discritised on $N$ non-uniformly
spaced points in radius, with higher resolution near the wall, and with
Fourier modes with index $|m|<M$ and $|k|<K$ in $\theta$ and $z$ respectively.
Our resolution is $(N,M,K)=(64,12,18)$, so that following the $\frac{3}{2}$-rule, variables are evaluated on $64\times36\times54$ grid points,
$(\Delta\theta\,D/2)^+\approx5$ and $\Delta z^+\approx7$.
Whilst this domain is
small, it is sufficiently large to reproduce the wall friction observed
for the infinite domain to within 10\%, and already sufficiently large to
exhibit a complex array of \po s (see \cite{WiShCv15} for details.)

\subsection{Symmetries of the pipe flow}
	\label{s:symmPipe}

Here we briefly review  the symmetries of the problem, and then focus on
the properties of the \emph{shift-and-reflect} flow-invariant subspace, to
which we restrict the study that we present in this article. For a
detailed discussion of flow-invariant subspaces of the pipe flow see,
for example, the Appendix of \cite{ACHKW11}.  It will be seen
presently that the \emph{shift-and-reflect} symmetry leads to two dynamically
equivalent regions of {\statesp}, later observed in simulations.

\begin{figure}
\centering
\setlength{\unitlength}{1.0\textwidth}
  \begin{picture}(1,0.25714286)%
    \put(0,0){\includegraphics[width=\unitlength]{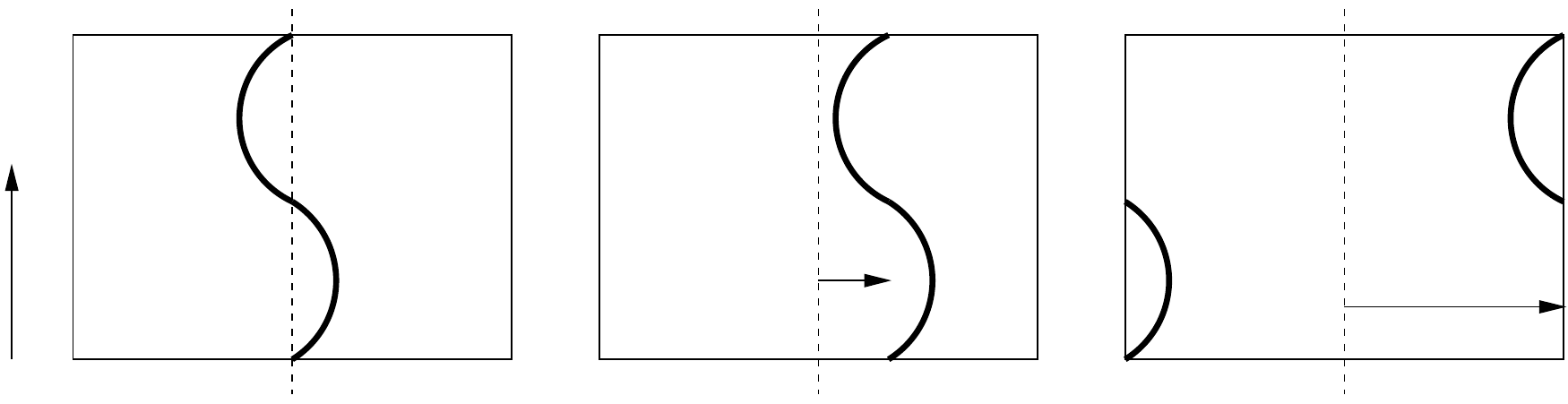}}%
    \put(0.00189372,0.16768848){\color[rgb]{0,0,0}\makebox(0,0)[lb]{\smash{$z$}}}%
    \put(0.43804,0.07252929){\color[rgb]{0,0,0}\makebox(0,0)[lb]{\smash{$\LieEl(\phi,0)$}}}%
    \put(0.87974908,0.07821044){\color[rgb]{0,0,0}\makebox(0,0)[lb]{\smash{$\LieEl_\theta$}}}%
    \put(0.15, 0.0){\color[rgb]{0,0,0}\makebox(0,0)[lb]{\smash{$(a)$}}}
    \put(0.487, 0.0){\color[rgb]{0,0,0}\makebox(0,0)[lb]{\smash{$(b)$}}}
    \put(0.823, 0.0){\color[rgb]{0,0,0}\makebox(0,0)[lb]{\smash{$(c)$}}}
  \end{picture}%
\\
\caption{\label{f-ShiftNRef}
(a) Sketch of a shift-and-reflect symmetric state in the doubly-periodic
domain $(\theta, z) \in [-\upi/m_0, \upi/m_0] \times [0, L]$ (same after
shift by $L/2$ in $z$ followed by reflection across the dotted line
$\theta=0$).
(b) In general, if the state in (a) is rotated by an angle $\phi$, then
the shift-and-reflect symmetry is broken.
(c) However, the state in (a) rotated by the half-shift $\phi=\upi/m_0$
remains in the shift-and-reflect subspace.
}
\end{figure}

In pipe flow the cylindrical wall restricts the rotation symmetry to
rotation about the $z$-axis, and translations along it. Let
$\LieEl(\gSpace,\shift)$ be the shift operator such that
$\LieEl(\gSpace,0)$ denotes an azimuthal rotation by $\gSpace$ about the
pipe axis, and $\LieEl(0,\shift)$ denotes the stream-wise translation by
$\shift$; let $\sigma$ denote reflection about the $\theta=0$ azimuthal
angle:
\bea
\LieEl(\gSpace,\shift) \, [u,v,w,p](r,\theta,z)
        & = & [u,v,w,p](r,\theta-\gSpace,z-\shift)
			  \continue
\sigma \, [u,v,w,p](r,\theta,z) \;\; & = & [u,-v,w,p](r,-\theta,z)
\,.
\label{pipeSymms}
\eea
The symmetry group of stream-wise periodic pipe flow is
$\SOn{2}_z \times \On{2}_\theta$;
in this paper we restrict our investigations to dynamics restricted to
the `shift-and-reflect' symmetry subspace 
\beq
   S = \{e,\sigma\LieEl_z\}
   \,,
\ee{ShiftReflOnl}
where $\LieEl_z$ denotes a streamwise shift by $L/2$,
\ie, flow fields \refeq{pipeSymms} that satisfy
\beq
	\lbrack u,v,w,p \rbrack(r,\theta,z)
=
	\lbrack u,-v,w,p \rbrack (r, - \theta,z - L/2)
    \,.
	\label{e-ShiftNRef}
\eeq
This requirement couples the stream-wise translations with the
azimuthal reflection. It is worth emphasising that by imposing the
symmetry $S$, \edit{continuous} rotations \edit{in $\theta$} are prohibited. Hence we consider only the
simplest example of a continuous group
\edit{for the stream-wise translations},
i.e.\ the one-parameter rotation group $\SOn{2}_z$,
omitting the subscript $z$ whenever that leads to no confusion.
In the azimuthal direction only a discrete rotation by
half the spanwise periodicity is allowed, i.e.\ by $\pi/m_0$.
We illustrate this property in \reffig{f-ShiftNRef}.
Thus, the symmetry
group $\Group$ of the pipe flow in the shift-and-reflect subspace
is
\beq
	\Group = \{\LieEl_{\theta} ,\,  \LieElz{l} \} \, ,
	\label{e-GSRpipe}
\eeq
where $\LieEl_{\theta}$ denotes the discrete azimuthal shift by $\pi/m_0$ and
$\LieElz{l} = \LieEl (0, l)$.

Solutions that can be mapped to each other by symmetry operations
\refeq{e-GSRpipe} are equivalent, \ie\ their physical
properties, such as instantaneous energy dissipation rates,
are same. Since the streamwise shift symmetry $\LieElz{l}$ is
continuous, one may have ``relative'' invariant solutions in the
\statesp\ of the pipe flow.
Such \cohStr s that we present in this paper are: (i) \Reqva\
\beq
	\vec{u}_{\mathrm{TW}} (\zeit) = \LieElz{c_{\mathrm{TW}} \,\zeit} \vec{u}_{\mathrm{TW}} (0) \, ,
	\label{e-TW}
\eeq
whose sole dynamics is a fixed velocity profile drifting along the axial direction with
constant phase speed $c_{\mathrm{TW}}$;
and (ii) \Rpo s
\beq
	\vec{u}_{\mathrm{RPO}} (\period{\mathrm{RPO}}) = \LieElz{l_{\mathrm{RPO}}} \, \vec{u}_{\mathrm{RPO}} (0) \, ,
	\label{e-RPO}
\eeq
which are time-varying velocity profiles which exactly repeat after
period $\period{\mathrm{RPO}}$, but shifted stream-wise by
$l_{\mathrm{RPO}}$.
In principle, one also has \rpo s
which are also relative with respect to azimuthal half rotations, such
that
\beq
	\vec{u}_{\mathrm{RPO}} (\period{\mathrm{RPO}}) = \LieEl_{\theta} \,
								\LieElz{l_{\mathrm{RPO}}} \,
								\vec{u}_{\mathrm{RPO}} (0) \, .
	\label{e-RPOtheta}
\eeq
These orbits connect two chaotic saddles
related by $\LieEl_{\theta}$. As we shall illustrate in the global
visualisations of dynamics in~\refsect{s:data}, such transitions
between the two saddles are quite rare. Therefore, we did not search for
\rpo s of \refeq{e-RPOtheta} kind, and focused instead on one of the
chaotic saddles related by azimuthal half-rotation.

\subsection{\StateDsp\ notation}
	\label{s:statespNot}

Let $\ssp$ denote the \statesp\ vector which uniquely represents a
three-dimensional velocity field $\vec{u}$ over the given computational
domain. While the \statesp\ representation $\ssp$ is technically
infinite-dimensional, due to numerical dicretization of the velocity
field (spatial discretization, truncated Fourier expansions, etc.) in
practice  $\ssp$  is a high- but always finite-dimensional vector.

We denote the semi-flow induced by the time evolution of the
\NSe\ \refeq{NavStokesDev} by $f^t$, so that
\beq
	\ssp (\zeit) = \flow{\zeit}{\ssp(0)}
	\label{e-flow}
\eeq
traces out a trajectory $\ssp (\zeit)$ in the \statesp.
For an infinitesimal time $\delta \zeit$, we can expand \refeq{e-flow}
as
$\ssp(\zeit+\delta \zeit)
= a(\zeit) + \vel (\ssp (\zeit)) \delta \zeit
+\mathcal O(|\delta\zeit|^2) $,
where
we refer to
\beq
	\dot{\ssp} = \vel (\ssp) \label{e-ODE}
\eeq
as the \emph{\stateDsp\ velocity}.

The ordinary differential equation (ODE) \refeq{e-ODE} has the same
symmetry group \refeq{e-GSRpipe} as the \NSe\ it approximates, \ie, the
\stateDsp\ velocity $\vel (\ssp)$ and the flow $\flow{\zeit}{\ssp}$
commute with the symmetry group actions,
$\LieEl \vel (\ssp) = \vel (\LieEl \ssp)$
and
$\LieEl \flow{\zeit}{\ssp} = \flow{\zeit}{\LieEl \ssp}$.

\section{Symmetry reduction by the \mslices}
\label{s:symmRed}

In this paper we investigate the geometry of turbulent attractor in terms
of shapes and unstable manifolds of a large number of \cohStr s that form
its backbone, and for that task a symmetry reduction scheme is absolutely
essential.
We recapitulate here briefly the construction of a symmetry-reduced
\statesp, or `slice'. For further detail and historical
notes the reader is referred to \refref{DasBuch}.

\begin{figure}
\centering
\includegraphics[width=.48\textwidth]{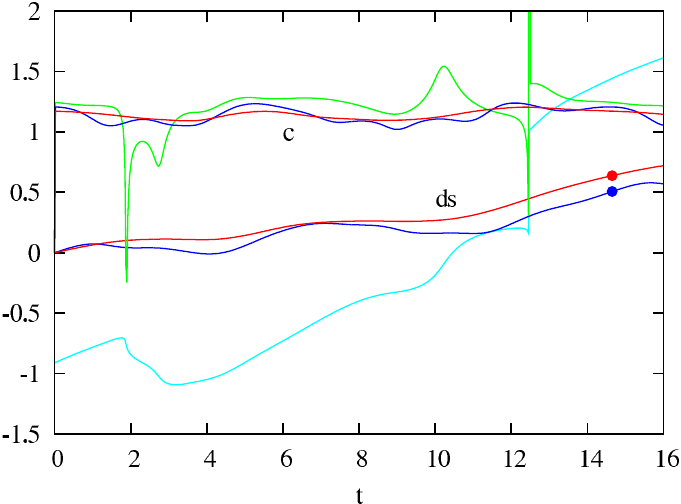}
\null\vspace{-25mm}
\includegraphics[width=.48\textwidth]{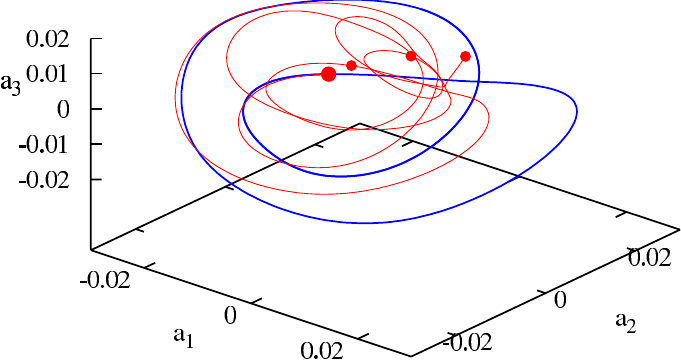}
\null\vspace{25mm}
\\
({\it a})\hspace{50mm}({\it b})
\caption{\label{fig:Sz}
    (Colour online)
Visualisations of the \rpo\ $\RPORUN{M/14.646}{9001}$.
({\it a})
Phase speed $c=\dot{\shift}$ and deviation in shift $ds=\shift-\bar{c}t$
from a Galilean frame moving with mean phase speed $\bar{c}=1.11$:
(green/cyan) $\slicep\equiv$\REQV{}{1.85}, (blue) template $\slicep$ on
$\RPORUN{M/14.646}{9001}$, taken to be the state for which the wall friction is lowest,
(red) method of connections.~
({\it b})
Three $\RPORUN{M/14.646}{9001}$ \edit{periods}, starting from the
large red dot, (blue) sliced with the template  $\slicep$, and (red)
method of connections, with times $\period{}$, $2\period{}$, $3\period{}$
marked by the small red dot. The projection on
${\ssp}_i=\langle\sspRed | e_i\rangle$, where the $e_i$ are unit vectors (see
main text).
}
\end{figure}

The set of points generated by
action of all shifts $\LieEl(\shift)$ on the \statesp\ point $\ssp$,
\beq
\pS_{\ssp } =
\{\LieEl(\shift) \, \ssp  | \, \shift \in [0, L) \}
\,,
\label{e-gOrbit}
\eeq
is known as the  {\em group orbit} of $\ssp$. All states in a group orbit
are physically equivalent, and one would like to construct a `symmetry-reduced
\statesp' where the whole orbit is represented by a single point $\sspRed$.
The {\em \mslices} accomplishes this in open neighbourhoods (never
globally), by fixing the shift $\shift$ with reference to a `template', a
\statesp\ point denoted $\slicep$.  A point on the group orbit with a
minimal distance from the template satisfies
\beq
0
\,=\,\frac{\partial}{\partial \shift}\,
   || \LieEl(-\shift)\,\ssp  - \slicep ||^2
\,=\,\frac{\partial}{\partial \shift}\,
   || \ssp  - \LieEl(\shift)\,\slicep ||^2
\,=\, 2\,
\inprod{\ssp  - \LieEl(\shift)\,\slicep }
    {- \,\frac{\partial}{\partial \shift} \LieEl(\shift)\,\slicep}
\label{eq:slicederiv}
\eeq
for a given $\shift$. Here, $\inprod{\cdot}{\cdot}$ denotes an
inner product and $\|\cdot\|$ denotes the corresponding norm
(see~\refsect{s:norms} for several specific choices of such inner products).
Let $\vec{t}'$ be the tangent to the group orbit of $\slicep$, \ie,
$\vec{t}'=\lim_{\delta\shift\to0}(\LieEl(\delta\shift)\,\slicep-\slicep)/\delta\shift$.
\edit{
Given that $\inprod{\slicep}{\slicep}$ is a constant,
\beq
 0=\frac{\partial}{\partial \shift} \inprod{\slicep}{\slicep}
   = 2\,\inprod{\slicep}{\frac{\partial}{\partial \shift} \slicep}
   = 2\,\inprod{\slicep}{\vec{t}'} .
\eeq
Using also that $\LieEl(\shift)$ and $\partial/\partial \shift$ commute,
then from \refeq{eq:slicederiv}
}
the minimum distance between the group orbit of $\ssp$ and the template
$\slicep$ occurs for a shift $\shift$ that
satisfies the{ \em slice condition},
\beq
   0    \,= \,
\inprod{\ssp  - \LieEl(\shift)\,\slicep}
       {\LieEl(\shift)\,\vec{t}'}
   \,= \,
\inprod{ \LieEl(-\shift)\,\ssp  - \slicep}{\vec{t}'}
   \,= \,
\inprod{\LieEl(-\shift)\,\ssp }{\vec{t}'}
\,.
\label{eq:slicecond}
\eeq
We denote by
$\sspRed=\LieEl(-\shift)\,\ssp$ the in-slice representative \edit{for
the whole} group orbit of the full \statesp\ state $\ssp$.
The in-slice trajectory $\sspRed(t)$ can be generated by integrating the
dynamics confined to the symmetry-reduced \statesp, or `slice',
\bea
\velRed (\sspRed) &=& \vel(\sspRed)
- \dot{\shift}(\sspRed) \, \vec{t}(\sspRed)
\,,\label{eq:slice}\\
\dot{\shift} (\sspRed) &=&
\inprod{ \vel (\sspRed)}{ \vec{t}'}
\, / \, \inprod{\vec{t}(\sspRed)}{ \vec{t}'}
\,.
\label{eq:reconst}
\eea
\edit{The first of these two equations expresses how the
symmetry-reduced {\stateDsp} velocity differs from the full {\stateDsp}
velocity by a small shift along the group orbit, parallel to
the tangent, at each instant in time.
Taking the inner product with $\vec{t}'$
leads to the second equation for $\dot\shift(t)$.
}
In a time-stepping scheme one has a good estimate for $\shift(t)$ from
its previous time step value, so it is more practical to use the slice
condition (\ref{eq:slicecond}) rather than (\ref{eq:reconst}) to
determine $\shift$. The latter condition, however, known as the {\em
reconstruction equation}, is useful in illustrating the behaviour of the
phase speed $\dot{\shift}$ (for an example, see \reffig{fig:Sz}\,({\it
a})). When the symmetry-reduced state $\sspRed$ and the template $\slicep$
are not too distant, their group tangents are partially aligned, and the
divisor in (\ref{eq:reconst}) is positive. If the tangents become
orthogonal, a division by zero occurs, and the phase speed
diverges. This defines the \emph{slice border}.

\refFig{fig:Sz}\,({\it a}) shows the \mslices\ applied to the \rpo\
$\RPORUN{M/14.646}{9001}$ (see \reftab{t-data}). For purposes of illustrating
that the slice hyperplane defined by (\ref{eq:slicecond}) is good only in an
open neighborhood, we take first a point on the somewhat distant \reqv\
$\TWRUN{1.845}{6416}$ as a trial template $\slicep$.  At time $\zeit=2$,
$\sspRed(\zeit)$ approaches the slice border, with a rapid change in
$\shift(\zeit)$ and large $c=\dot{\shift}$.  Near time $\zeit=12$ the orbit
hits the border, and there is a discontinuity in $\shift$.

A nearer template point would be a better choice. Indeed, as illustrated by
\reffig{fig:Sz}\,({\it b}), we find that the point of the lowest wall
friction on the orbit itself works very well as a template state $\slicep$:
the slice now captures the entire $\RPORUN{M/14.646}{9001}$ without encountering any
slice border and any discontinuity (blue).

\subsection{Method of connections}
	\label{s:statespNot}
If one were to use the state itself as a template at each instant in time,
then a division by zero would always be avoided.  This corresponds to
projecting out the component parallel to the shift, \ie, the group orbit
tangent, at each moment in time.
In \refref{rowley_reconstruction_2000} this is referred to as the {\em method of
connections}. While very appealing and sometimes deployed \citep{KrZaEc14}
to `calm' a turbulent flow,
the {method of connections} is \emph{not} a symmetry reduction method, in the
sense that the dimensionality of the \statesp\ is not reduced by 1 for each
continuous symmetry parameter
\edit{(see, e.g., \refref{rowley_reduction_2003,atlas12})}.
This is illustrated by \reffig{fig:Sz}. Starting with the same initial
condition (fat red point), with the \mslices\ the orbit
$\sspRed(0)=\sspRed(T)$ closes after one period $T=14.646$  (blue), while with
the method of connections the orbit continues filling out the \rpo\ torus
ergodically, never closing into a \po.
\refFig{fig:Sz}\,({\it b}) shows a projection of the trajectory generated
by the two approaches, for three cycles of the \po.
Whereas the orbit closes for the \mslices, for the
{method of connections} the invariant torus remains a torus.
\edit{
In conclusion, nothing is gained by using the method of connections.
    }

The orthogonal unit vectors used in \reffig{fig:Sz}({\it b}) are the
leading components from a \edit{principal component analysis (PCA),}
using 586 symmetry-reduced states
equi-spaced in time around the orbit.  Further details are given
in~\refsect{s:stateSp}.

\subsection{\FFslice}
\label{s:fFslice}

The \mslices\ is local, and the slice border discontinuity is avoided only within
a neighbourhood of the template. In our previous work
\citep{ACHKW11}, a switching approach was applied to help ensure closeness
to the template.  Whilst this enabled symmetry reduction of longer
trajectories, it proved difficult to switch before reaching a border
whilst simultaneously ensuring continuity of $\shift(t)$.

A different approach was taken by \refref{BudCvi14} for one-dimensional PDEs with \SOn{2}
symmetry, where it was shown that the first term in the Fourier expansion of
the flow field can be used as the template for a \emph{global} slice, border
of which is never visited by generic ergodic trajectories. This method relied
on the observation that projections of group orbits on the subspace spanned by
the first Fourier mode components (sine and cosine) are non-overlapping
circles; hence one can find a unique polar angle in this projection to
quotient the \SOn{2} symmetry. For a scalar field
$u(x, \zeit) = u(x + L, \zeit)$ in one periodic space dimension,
a slice template of the form
$u' = a \cos (2 \pi x / L) + b \cos (2 \pi x / L)$, where $a$ and
$b$ are constants, defines a \fFslice .
In higher dimensions, one has more freedom in choosing \fFslice\
templates. For pipe flow, any $\slicep$ corresponding to a velocity
field of the form
\beq
	\vec{u}' (r, \theta, z) =
	\vec{u}_c(r, \theta) \cos (\alpha z) +
	\vec{u}_s(r, \theta) \sin (\alpha z)
	\, ,
	\label{e-ffSliceTemp}
\eeq
where $\vec{u}_{c,s}$ are three-dimensional vector fields that depend only
on $r$ and $\theta$, can be a candidate for a \fFslice\ template.
The vector fields
$\vec{u}_{c,s}$ should be chosen such that the slice border condition
\(
	\inprod{\ssp}{\slicep}
\,+\,
    \mathrm{i}\inprod{\ssp}{\LieElz{L/4} \slicep}
        = 0
\)
is avoided by generic flow fields
$\vec{u}= \vec{u} (r, \theta, z, \zeit)$.

All slices are local, but since
\(
\inprod{\LieEl(-\shift)\,\ssp}{\vec{t}'}
=\inprod{\ssp}{\LieEl(\shift)\,\vec{t}'}
\,,\)
one way to construct slice hyperplanes with larger domains of validity is by
picking templates with smoother group orbits. Smoother states (\ie, states
dominated by low Fourier modes) tend to be associated with lower dissipation
or wall friction.
Guided by this intuition, we construct the \fFslice\ template
\refeq{e-ffSliceTemp} by taking a low-dissipation solution from the turbulent
set and setting all of its components to $0$ other than the ones with axial
Fourier modes $k=1$ \citep{WiShCv15}.
For the calculation in \reffig{fig:Sz}, such a template
was capable of capturing the whole orbit.

The slice-fixing shift $\shift (\zeit)$ for a trajectory $\ssp (\zeit)$ is
computed from the polar angle in the plane spanned by
$(\slicep, \LieElz{L/4} \slicep)$ as
\beq
	\shift (\zeit)  = \frac{L}{2 \pi }\,
					  \mbox{Arg}\left[  \inprod{\ssp (\zeit)}{\slicep}
					  + \mathrm{i} \inprod{\ssp (\zeit)}{
								  \LieElz{L/4} \slicep}\right],
	\label{e-fFsliceShift}
\eeq
where $\mbox{Arg}$ denotes the argument of the complex number.
One can then find the translation symmetry-reduced trajectory $\hat{\ssp}
(\zeit)$ by shifting the full \statesp\ trajectory $\ssp (\zeit)$ back to
slice by
\(
  	\hat{\ssp} (\zeit) = \LieElz{- \shift(\zeit)} \ssp (\zeit)
\,.
\)

\section{\CohStr s}
\label{s:data}

By carrying out an extensive search for \reqva\ \refeq{e-TW} and \rpo s
\refeq{e-RPO}, we have found  $8$ \reqva\ and \NRPOsTot\ \rpo s of the
pipe flow, listed in \reftab{t-data}.
The \reqva\ are labelled by their mean dissipation $\bar{D}$ (in units of
the kinetic energy $E_{0}$ of the laminar solution), and \rpo s by
their periods $T$ (in units of $D/U$).
The numerical method for finding most of these invariant solutions is the
Newton--GMRES--hook iteration, discussed in detail in \refref{Visw07b}
and \refref{ChaKer12}.
\Rpo s with long periods tend to be more difficult (or impossible) to
find with the standard Newton--GMRES--hook iteration. In order to capture
such long orbits, we implemented a multiple-shooting Newton method,
outlined in \refappe{s:multi}, and found $5$ \rpo s, marked with
subscript `M' in \reftab{t-data}.
The highly symmetric
$N4$ type \reqv\ of \refref{Pringle09} belongs to an invariant subspace
with an additional shift-and-rotate symmetry,
\beq
\lbrack u,v,w,p \rbrack(r,\theta,z)
=
\lbrack u,-v,w,p \rbrack (r,\theta - \pi / m_0,z - L/2)
\,.
\label{e-ShiftNRot}
\eeq
The \reqva\ $N4$  appear as a lower/upper branch pair and are therefore
labeled as $N4L$ (lower branch) and $N4U$ (upper branch). The terminology
refers to the appearance of these solutions from a saddle node
bifurcation at a lower \Reynolds\ number as a pair of solutions with
low ($N4L$) and high ($N4U$) dissipation rates. The lower branch
solution is believed to belong to the laminar-turbulent boundary
\citep{Pringle09}.

\begin{table}
	\begin{tabular}{l c c c c c | l c c c c c}
Solution & $\bar{D}$ & $\bar{c}$ & $d_U$ & $\mu^{\max}$ & $\omega$ or $\theta$ & Solution & $\bar{D}$ & $\bar{c}$ & $d_U$ & $\mu^{\max}$ & $\omega$ or $\theta$ \\
\hline\\
$\TWRUN{N4L/1.38}{6481}$\ddag & $1.38$ & $1.238$ & $3$ & $0.1809$ & $0.0$ & $\TWRUN{1.578}{6492}$ & $1.578$ & $1.108$ & $9$ & $0.2877$ & $0.0$ \\
$\TWRUN{2.039}{6494}$\ddag & $2.039$ & $1.091$ & $7$ & $0.1159$ & $0.0$ & $\TWRUN{1.845}{6416}$ & $1.845$ & $1.039$ & $11$ & $0.5166$ & $0.891$ \\
$\TWRUN{1.968}{6472}$\ddag & $1.968$ & $1.105$ & $9$ & $0.1549$ & $0.259$ & $\TWRUN{1.783}{6459}$ & $1.783$ & $1.035$ & $8$ & $0.323$ & $1.119$ \\
$\TWRUN{1.885}{6491}$ & $1.885$ & $1.073$ & $8$ & $0.4568$ & $0.206$ & $\TWRUN{2.041}{6480}$\ddag & $2.041$ & $1.095$ & $8$ & $0.1608$ & $0.0$ \\
$\TWRUN{N4U/3.28}{6482}$\ddag & $3.279$ & $1.051$ & $30$ & $0.9932$ & $1.89$ & $\TWRUN{1.926}{6495}$ & $1.926$ & $1.096$ & $8$ & $0.2504$ & $0.414$ \\
\hline\\
$\RPORUN{F/6.668}{6669}$\ddag & $1.805$ & $1.12$ & $3$ & $0.0534$ & $1.69$ & $\RPORUN{F/M/33.81}{9003}$ & $1.805$ & $1.128$ & $5$ & $0.0471$ & $1.727$ \\
$\RPORUN{F/13.195}{6652}$\ddag & $1.839$ & $1.117$ & $5$ & $0.0581$ & $2.038$ & $\RPORUN{F/33.968}{6658}$ & $1.806$ & $1.127$ & $5$ & $0.0588$ & $1.671$ \\
$\RPORUN{F/20.427}{6465}$\ddag & $1.809$ & $1.128$ & $5$ & $0.0771$ & $0.0$ & $\RPORUN{F/40.609}{9011}$ & $1.814$ & $1.125$ & $5$ & $0.0505$ & $0.315$ \\
$\RPORUN{F/26.861}{9005}$ & $1.84$ & $1.121$ & $5$ & $0.0679$ & $\pi$ & $\RPORUN{F/M/47.449}{9004}$ & $1.826$ & $1.126$ & $5$ & $0.0586$ & $\pi$ \\
$\RPORUN{F/26.964}{6654}$ & $1.826$ & $1.124$ & $6$ & $0.0493$ & $0.986$ & $\RPORUN{F/M/53.876}{9006}$ & $1.83$ & $1.124$ & $6$ & $0.0457$ & $1.253$ \\
$\RPORUN{F/27.299}{6655}$\ddag & $1.815$ & $1.126$ & $4$ & $0.0678$ & $0.961$ &  \edit{$\RPORUN{F/M/67.936}{9012}$} & 
\edit{$1.806$} & \edit{$1.128$} & \edit{$5$} & \edit{$0.0587$} & \edit{$2.945$} \\
\hline\\
$\RPORUN{4.954}{8100}$\ddag & $2.015$ & $1.084$ & $3$ & $0.1509$ & $1.643$ & $\RPORUN{M/14.544}{9010}$ & $2.015$ & $1.102$ & $6$ & $0.1846$ & $0.0$ \\
$\RPORUN{5.468}{8189}$ & $2.003$ & $1.091$ & $6$ & $0.1452$ & $1.351$ & $\RPORUN{M/14.646}{9001}$ & $1.776$ & $1.133$ & $5$ & $0.1473$ & $\pi$ \\
$\RPORUN{6.119}{8192}$ & $1.875$ & $1.081$ & $7$ & $0.1912$ & $0.0$ & $\RPORUN{14.961}{5001}$ & $1.945$ & $1.114$ & $5$ & $0.1915$ & $0.878$ \\
$\RPORUN{6.134}{6606}$ & $1.86$ & $1.086$ & $7$ & $0.1596$ & $0.0$ & $\RPORUN{15.081}{6479}$ & $2.06$ & $1.081$ & $8$ & $0.1392$ & $0.0$ \\
$\RPORUN{6.18}{8191}$ & $1.865$ & $1.091$ & $5$ & $0.211$ & $0.0$ & $\RPORUN{15.46}{6607}$\ddag & $1.781$ & $1.146$ & $7$ & $0.1166$ & $0.0$ \\
$\RPORUN{6.359}{8194}$ & $1.769$ & $1.054$ & $11$ & $0.2614$ & $0.0$ & $\RPORUN{15.798}{8082}$ & $1.869$ & $1.125$ & $6$ & $0.1089$ & $\pi$ \\
$\RPORUN{6.458}{8183}$ & $2.117$ & $1.074$ & $7$ & $0.2055$ & $0.0$ & $\RPORUN{15.915}{8154}$ & $1.951$ & $1.106$ & $8$ & $0.1547$ & $\pi$ \\
$\RPORUN{7.246}{8146}$ & $1.982$ & $1.105$ & $5$ & $0.209$ & $0.0$ & $\RPORUN{15.972}{8206}$ & $1.956$ & $1.097$ & $7$ & $0.1473$ & $\pi$ \\
$\RPORUN{7.272}{8016}$ & $2.015$ & $1.1$ & $5$ & $0.1852$ & $0.0$ & $\RPORUN{16.271}{6356}$ & $1.978$ & $1.09$ & $7$ & $0.1454$ & $1.977$ \\
$\RPORUN{7.423}{8024}$\ddag & $1.838$ & $1.109$ & $6$ & $0.1195$ & $0.387$ & $\RPORUN{16.878}{6422}$ & $1.969$ & $1.099$ & $5$ & $0.1219$ & $\pi$ \\
$\RPORUN{7.741}{8058}$\ddag & $1.707$ & $1.138$ & $5$ & $0.0983$ & $0.0$ & $\RPORUN{17.21}{8112}$ & $1.999$ & $1.098$ & $7$ & $0.1523$ & $\pi$ \\
$\RPORUN{9.735}{8027}$\ddag & $2.05$ & $1.086$ & $7$ & $0.1872$ & $\pi$ & $\RPORUN{17.46}{8142}$\ddag & $1.917$ & $1.121$ & $6$ & $0.0842$ & $0.205$ \\
$\RPORUN{11.696}{8023}$ & $1.961$ & $1.108$ & $9$ & $0.1129$ & $\pi$ & $\RPORUN{21.704}{8108}$ & $1.868$ & $1.12$ & $7$ & $0.0951$ & $\pi$ \\
$\RPORUN{12.026}{8095}$ & $2.09$ & $1.088$ & $6$ & $0.1476$ & $0.0$ & $\RPORUN{22.063}{8122}$ & $2.032$ & $1.101$ & $7$ & $0.1352$ & $1.723$ \\
$\RPORUN{12.566}{6351}$ & $2.053$ & $1.083$ & $10$ & $0.1677$ & $\pi$ & $\RPORUN{23.047}{8029}$ & $1.874$ & $1.12$ & $6$ & $0.1848$ & $0.0$ \\
$\RPORUN{12.706}{8214}$ & $2.156$ & $1.07$ & $6$ & $0.1692$ & $1.083$ & $\RPORUN{23.356}{8087}$\ddag & $1.98$ & $1.112$ & $6$ & $0.101$ & $1.249$ \\
$\RPORUN{13.592}{6475}$ & $1.987$ & $1.099$ & $7$ & $0.1072$ & $0.0$ & $\RPORUN{26.049}{6426}$ & $2.028$ & $1.097$ & $8$ & $0.1635$ & $\pi$ \\
$\RPORUN{14.045}{6477}$\ddag & $1.903$ & $1.107$ & $6$ & $0.1403$ & $\pi$ & $\RPORUN{27.238}{5003}$ & $1.992$ & $1.098$ & $8$ & $0.1258$ & $0.0$ \\
	\end{tabular}
	\caption{The list of the \cohStr s reported in this work.
        Average rate of dissipation $\bar{D}$, average down-stream phase
        velocity $\bar{c}$, dimension of the unstable manifold $d_U$,
        real part of the largest stability eigenvalue / Floquet exponent
        $\mu^{max}$ is shown. Last column corresponds to the imaginary
        part $\omega$ of the leading stability eigenvalue for {\reqva},
        and phase $\theta$ of the leading Floquet multiplier for \rpo s.
        {\Reqva} are labeled by their dissipation rate $\bar{D}$,
        \rpo s by their period $\period{}$. A family of twelve \rpo s
        which appear to have similar physical properties are grouped
        together and labeled with subscript $F$.
        \edit{
        Solutions marked with \ddag \; were previously reported in Table
        1 of \refref{WiShCv15}.
        The six solutions marked with $M$ were obtained by the
        multiple-shooting Newton method of \refappe{s:multi}.
             }
        }
	\label{t-data}
\end{table}

The \reqva\ linear stability exponents $\lambda_j = \mu_j + i \omega_j$
are computed by linearizing the governing equations in the co-moving
frame, in which a \reqv\ becomes an \eqv.
The leading stability exponent, \ie, the exponent with the largest real part,
is reported in \reftab{t-data}.
The integer $d_U$ denotes the number of exponents with positive real
parts, $\mu_j > 0$, which determine the dimension of the unstable
manifold of the \reqv.

The linear stability of a \rpo\ is described by its Floquet multipliers
$\Lambda_j = \exp(\mu_j T + i \theta_j)$.
As for the \reqva, in \reftab{t-data} we report the number of the
unstable directions $d_U$, real part $\mu^{max}$ of the leading Floquet
exponent
 $\lambda_j = (1/T) \ln |\Lambda_j|$,
and the phase $\theta$ of the leading Floquet multiplier of the \rpo.

Twelve \rpo s listed in \reftab{t-data} (indicated by subscript $F$) are
separated from the rest in the table. We refer to these orbits as the
\emph{first family}  solutions, due to their remarkably similar physical
and dynamical properties. The periods of the first family members are
approximately integer multiples of the shortest \rpo, whose period is
$\period{}=6.668$.
Indeed, numerical continuations in $Re$ and/or geometry parameters show
that several first family members originated from bifurcations off the
parent orbit \RPORUN{F/6.668}{6669}.
For example, \RPORUN{F/13.195}{6652} is born out of a period-doubling
bifurcation at $Re=2191$. As is shown in the next section, the
first family orbits lie near each other in all \statesp\ visualisations,
populating a small region of the \statesp. A detailed bifurcation
analysis of the first family orbits is the subject of ongoing research
\citep{ShWi16}.

\section{{\StateDsp} visualisation of fluid flows}
\label{s:stateSp}

With the available computational resources, today one can generate a large number of
turbulent trajectories as solutions of the \NSe\ with various initial conditions.
What can one learn from the resulting enormous amounts of data?

A routine approach is to seek to understand the statistical properties of
physically relevant quantities such as velocity correlations, enstrophy,
palinstrophy, etc. One objective of the program of determining
\emph{\cohStr s} is to go beyond a statistical description, and explore
the {\statesp} \emph{geometry} of long-time attractors of such dissipative
flows. This should ultimately provide a coarse-grained partition of the
{\statesp} into regions of qualitatively and quantitatively similar
behaviours.

Embarking on this path, one is immediately confronted with several
fundamental dilemmas with no known resolution:
\begin{enumerate}
\item \emph{\Statesp\ geometry.}
    Inertial manifolds and attracting sets of nonlinear dissipative flows
    are nonlinear, curved subsets of the full \statesp. Even for the
    H\'enon attractor we only have a partial understanding of the
    topology \citep{pre88top,dCH02} and the existence of such attracting
    sets \citep{BenCar91}. Our strategy for visualisation of the
    `{\statesp} geometry' of the \NSe\ is to populate it by {\cohStr
    s}, e.g. \eqva, {\reqva}, \po s and \rpo s, and capture the local
    ``curvature'' of the attractor by tracing out segments of their
    unstable manifolds and their heteroclinic connections (see, e.g.,
    \refref{GHCW07,GHCV08})
\item \emph{Measuring distances.}
    The distance between two fluid states is measured using some norm.
    There is no solid physical or mathematical justification for using
    the usual $L_2$ or `energy' norm. For example, in some problems a
    Sobolev norm might be preferred in order to either penalize or
    emphasize the small scale structures (see, e.g.,
    \refref{Mathew07,Lin2011,Faraz15}
    \edit{
    and \refsect{s:TW-norm}).
    Furthermore, in presence of continuous and discrete symmetries, it is
    absolutely imperative that symmetries be reduced before a distance can
    be measured (see, e.g., \refeq{eqProjMinDist}); states on group
    orbits of nearby states can lie arbitrarily far in the \statesp. As
    different choices of a slice yield different distances, this
    introduces a further arbitrariness into the notion of `distance'.
          }
\item \emph{Low-dimensional visualisations}. The \statesp\ of \NSe\ is
    infinite-dimensional. To visualize the geometry of the invariant
    solutions one inevitably projects the solutions to two- or
    three-dimensional subspaces.  For a discussion of optimal projections that best
    illuminate the structure of an attractor, see \refref{statespDummies}.
\end{enumerate}

Although we are in no position to resolve any of these issues in this
paper, we will elucidate, through examples, the impact of the choice one
makes in answering each question.

\subsection{Choice of the norm}
 \label{s:norms}

In this work, we use two rather different norms, the standard {energy norm},
and a hand-crafted `\lowpass' norm. In what follows, we show how the choice
of the norm can significantly alter the \statesp\ visualisations, and the
conclusions drawn from them.

Let  $\vec{u}=\sum_{km}\vec{u}_{km}(r)\exp(2\mathrm{i}\alpha
kz+\mathrm{i}m_0m\theta)$ denote the Fourier series of a velocity field
$\vec{u}$ defined in a pipe of axial length $L=\pi/\alpha$. The variables
$\vec{u}_{km}$ denote the Fourier coefficients corresponding to the axial and
azimuthal directions as functions of the radial distance $r$. For two
velocity fields $\vec{u}_1$ and $\vec{u}_2$, we define the $L^2$ inner
product
\bea
\inprodE{\vec{u}_1}{\vec{u}_2} &=& \frac{1}{2 E_\mathrm{HP}} \int_V
\vec{u}_1 \cdot \vec{u}_2 \, r\,d\theta \,dr \,d z
\label{e-inprodL2} \, ,\\
&=& \frac{1}{E_\mathrm{HP}} \int_0^{1/2} \!\! r \, dr \,
\sum_{k, m}
\vec{u}^*_{1,km}(r) \cdot \vec{u}_{2,km}(r) \, ,
\label{e-inprodL2Data}
\eea
where $V$ denotes the cylindrical flow domain and $E_\mathrm{HP}$ is the kinetic
energy of the Hagen-Poiseuille flow. In \refeq{e-inprodL2Data}, we
write the integral explicitly in terms of Fourier modes and radial
integration, which in practice are approximated numerically. This inner product
corresponds to the $L^2$ or the kinetic \emph{energy norm},
\beq
E(\vec{u}) = \frac{1}{2}\normE{\vec{u}}^2 = \frac{1}{2}\inprodE{\vec{u}}{\vec{u}}.
\label{e-normE}
\eeq

We sometimes find it more informative to use a metric that emphasizes larger
scale structures along the continuous symmetry directions. For this reason,
we define the \emph{`\lowpass'} metric,
\beq
\inprodLP{\vec{u}_1}{\vec{u}_2} = \frac{1}{V} \int_0^{1/2} r dr
\sum_{k, m}
\frac{1}{1 + (\alpha k)^2 + (m_0 m)^2} \,
\vec{u}^*_{1,km}(r) \cdot  \vec{u}_{2,km}(r)
\,,
\label{e-inprodLP}
\eeq
which penalizes higher Fourier modes (short wavelengths. In the axial and
azimuthal directions this is a variation of a Sobolev $H^{-1}$ norm
\citep{Lax02}: The weights are smaller for larger values of $k$ and $m$,
hence shorter wavelengths are de-emphasized.

In the work reported here, we rely primarily on the energy
norm~\eqref{e-inprodL2Data}, except for \refsect{s:TW-norm} where we
contrast {\statesp} visualisations using the energy and the
\lowpass\ norms, and comment on their relative effectiveness in our
searches for \rpo s.

\subsection{Global visualisations: Principal Component Analysis}
\label{s-global}

We begin our investigation of \statesp\ with a data-driven method
in order to obtain a general qualitative picture. The use of principal
component analysis (PCA), otherwise known as
proper orthogonal decomposition (POD) in the context of fluids,
has been well documented (see e.g.\ \cite{Berkooz93}).
Broadly speaking, the method extracts a set of orthogonal vectors that
span the data with minimal residual, with respect to some norm.

Here we apply the method to extract principal components, relative to the
mean \edit{$\bar{\sspRed}$} of the data set of $N$ states $\sspRed_i$
in the symmetry-reduced \statesp, symmetrized
with respect to $\LieEl_{\theta}$.
Singular value decomposition is applied to the matrix of inner products
of the deviations
$\tilde{\sspRed}_i=\sspRed_i-\bar{\sspRed}$,
\beq
R_{ij}=\frac{1}{N-1}
\inprod{\tilde{\sspRed}_i}{\tilde{\sspRed}_j}_\mathrm{L^2}\, ,
\qquad
R = U\,S\,V^T \, ,
\eeq
from which the $j^\mathrm{th}$ principal component is calculated
\beq
\vec{e}_j = \sum_{i=1}^N \tilde{\sspRed}_j U_{ij}\, ,
\qquad
\hat{\vec{e}}_j = \vec{e}_j /
\inprod{{\vec{e}}_j}{{\vec{e}}_j}_\mathrm{L^2} \, .
\ee{e-PCAprojection}
Each principal component $\hat{\vec{e}}_j$ has the property
that the root-mean-square of the projection
$p_j=\inprod{\tilde{\sspRed}_i(t)}{\hat{\vec{e}}_j}$ (taking the mean over $i$)
equals the $j^\mathrm{th}$ singular value, \edit{$S_{jj}$,} of the correlation matrix.

\begin{figure}
	\centering
	(a) \includegraphics[width=0.4\textwidth]{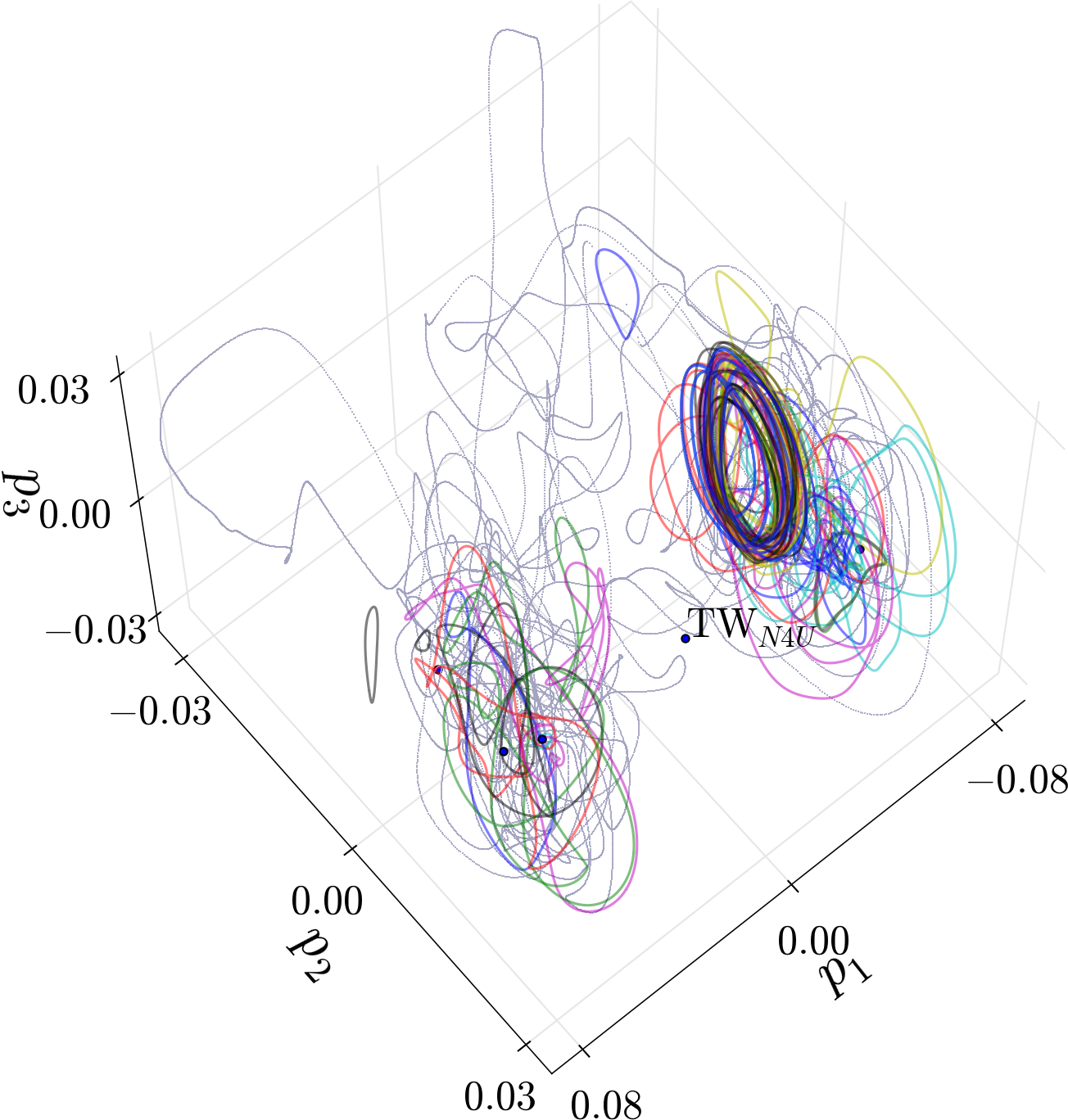}
	(b) \includegraphics[width=0.4\textwidth]{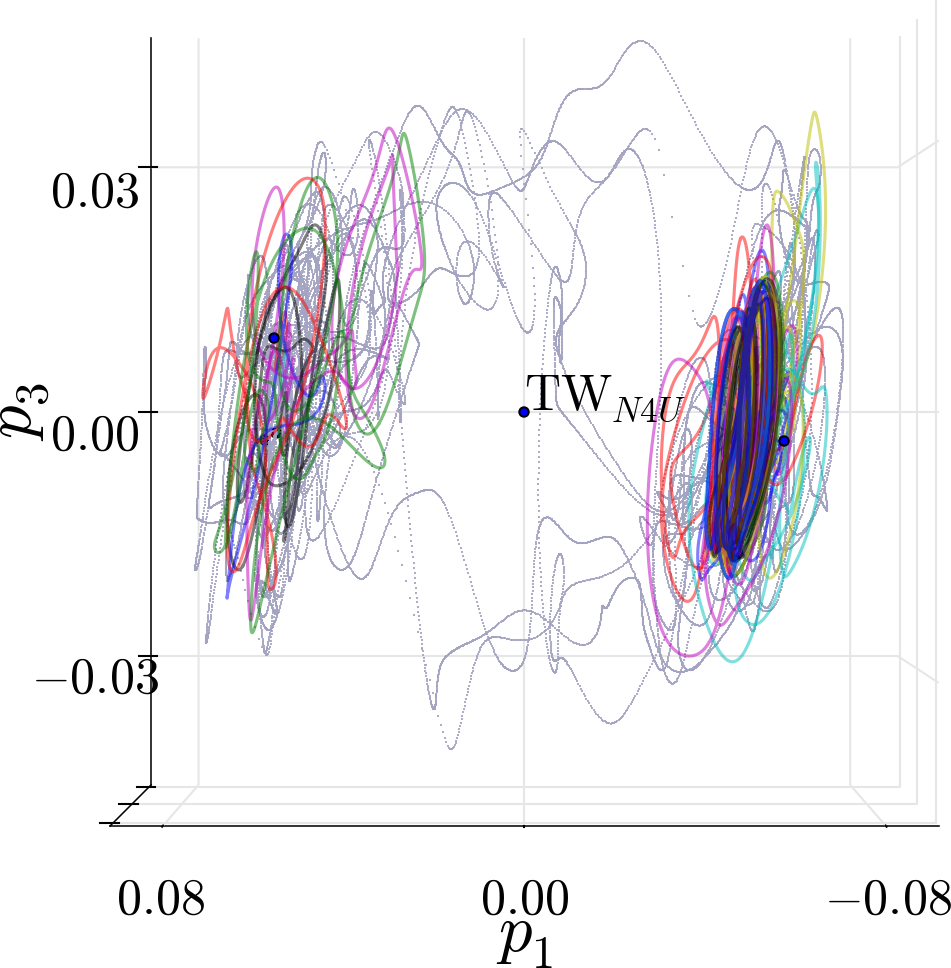}
	\caption[PCA projections of pipe flow solutions]{\label{f-PCAFull}
        (Colour online)
		\Rpo s (closed curves, different colors),
		$5$ {\reqva} (marked with black dots),
		and an two turbulent trajectories (gray, transparent dots)
		of the pipe flow projected onto first three principal
		components \edit{\refeq{e-PCAprojection}}, two different viewing angles.
	}
\end{figure}

\edit{Principal components were extracted from $2000$ uncorrelated
symmetry-reduced states obtained from ergodic trajectories.}
\refFig{f-PCAFull} shows
all our \rpo s, $5$ {\reqva},
and two turbulent trajectories projected onto the principal components,
computed as above. There are some notable observations
about \reffig{f-PCAFull}: Firstly, \po s appear to be
localized on two sides of the $p_1 = 0$ plane, and the turbulent
trajectories rarely switch from one side to other.
The \fFslice\ reduces the continuous
translation symmetry of pipe flow but the discrete half-rotation
symmetry $\LieEl_{\theta}$ still remains; the two
sides of \reffig{f-PCAFull} are related to each other by this discrete symmetry
operation. Also note that the highly symmetric
$N4$ {\reqv} $\TWRUN{N4U/3.28}{6482}$, which is
invariant under $\LieEl_{\theta}$, appears to lie at the origin of
$(p_1, p_3)$ plane.

It is clear from \reffig{f-PCAFull} that the principal component
$\hat{\vec{e}}'_1$ is aligned along the symmetry direction. This makes
the information contained along this direction redundant, since each
solution with $p_1 > 0$ has a copy with $p_1 < 0$.  As we are
most interested in the details of the turbulent set and invariant dynamical
behaviour of the system, our next step is to reduce the
discrete $\LieEl_{\theta}$-symmetry. For this purpose, we define the
`fundamental domain' \citep{DasBuch} as $p_1 > 0$ and bring all our
data from turbulence simulations and {\cohStr s} to this
half of the \statesp. With the desymmetrized turbulence data, we
recompute principal components $\tilde{\vec{e}}'_j$, which we will
refer to as `fundamental domain principal components'.

\begin{figure}
	\centering
	(a) \includegraphics[width=0.4\textwidth]{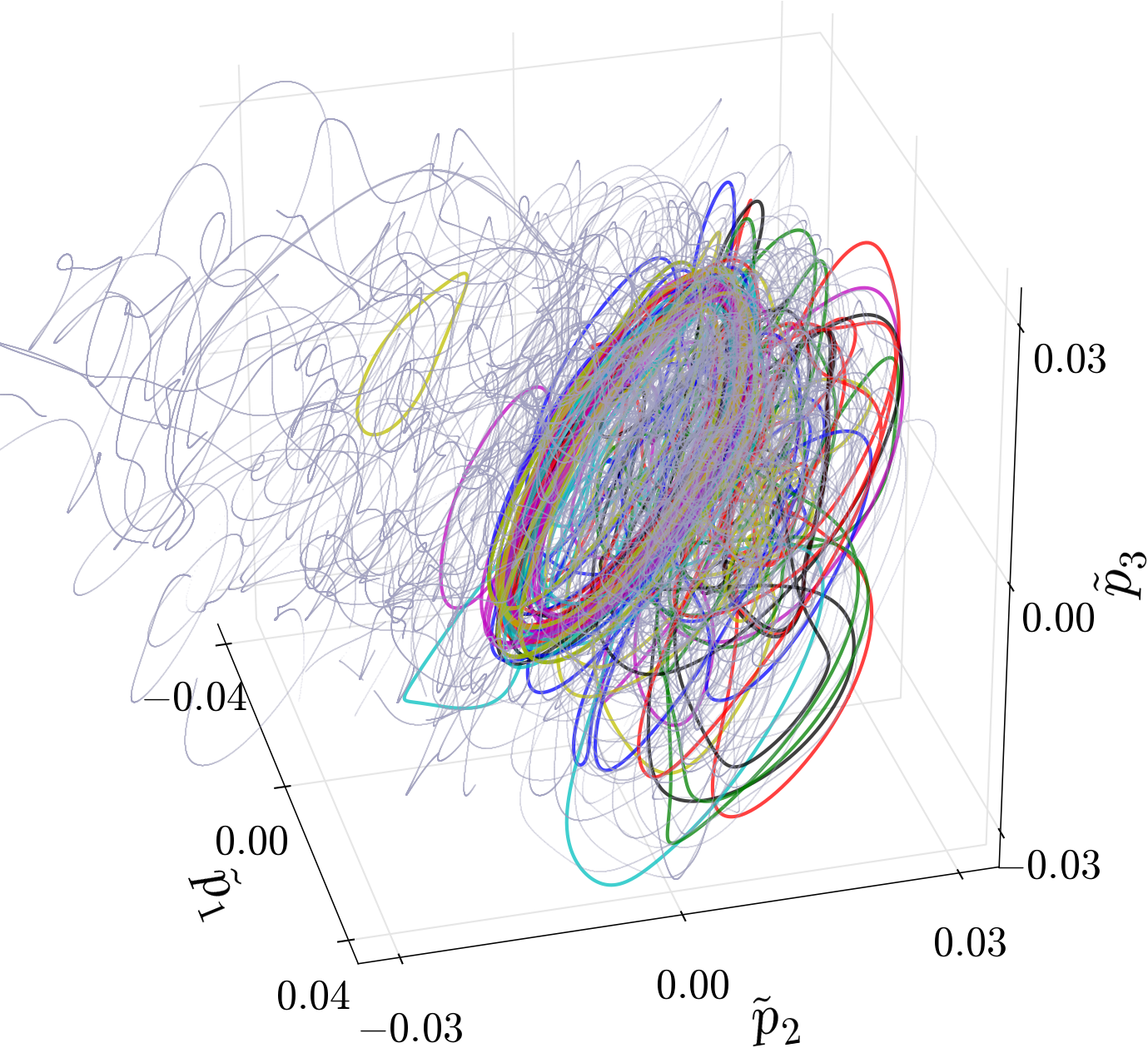}
	(b) \includegraphics[width=0.4\textwidth]{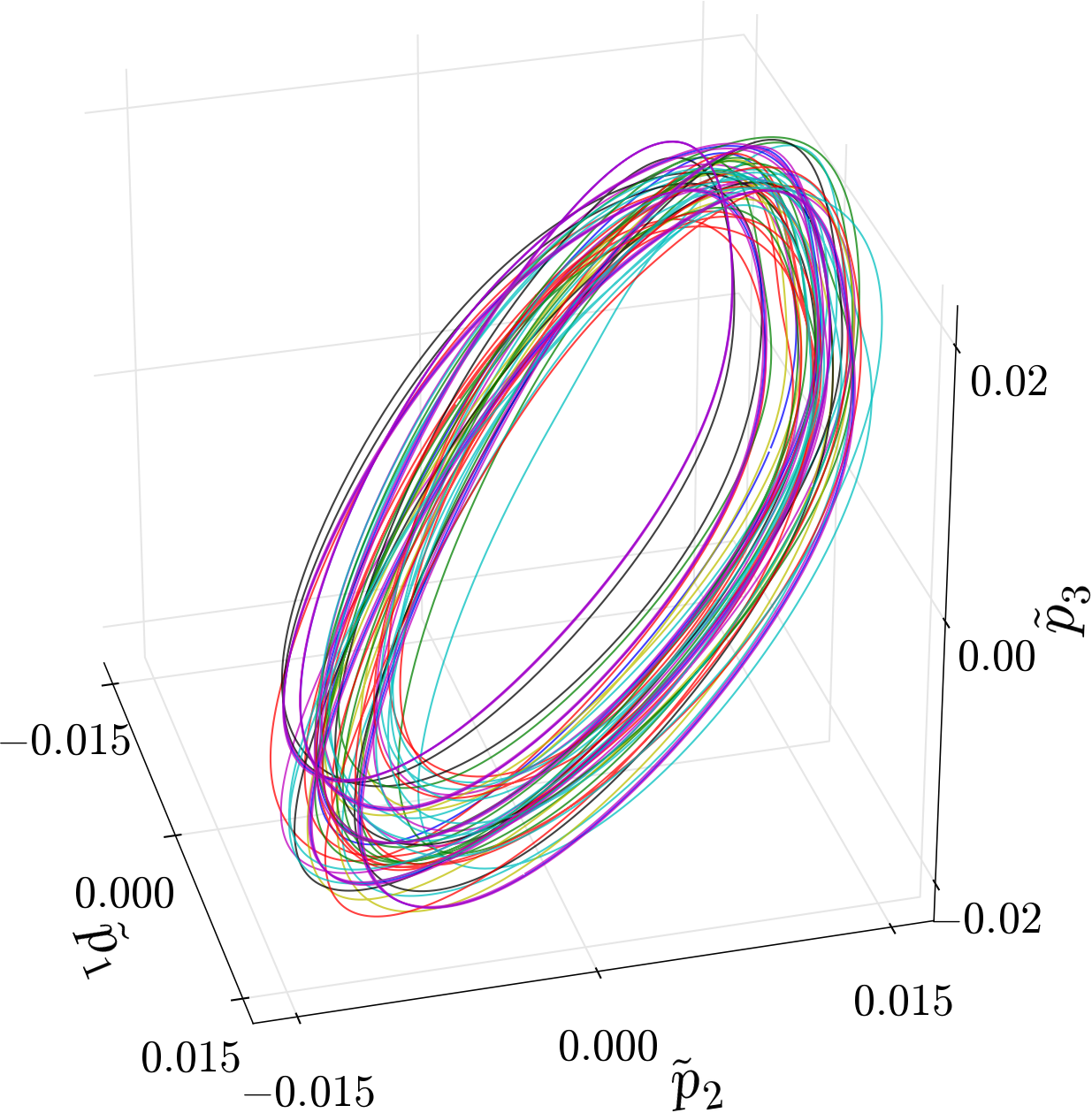}
	\caption[Pipe flow solutions in the fundamental domain]{
		\label{f-PCAFunD}
        (Colour online)
		(a) \Rpo s (closed curves of various colors) and
		5 turbulent trajectories (gray dots)
		projected onto first three fundamental domain
		principal components.
		(b) $11$ orbits, which appear to fill out a region of the
		\statesp.
	}
\end{figure}

\refFig{f-PCAFunD} shows \rpo s and turbulent trajectories projected onto
the three dominant fundamental domain principal components. In
\reffig{f-PCAFunD}, we no longer have two ``clouds'' and \rpo s are mostly
located in the region of \statesp\ where turbulent trajectories
spend most of their time.  One striking observation from
\reffig{f-PCAFunD}\,({\it a}) is that a subset of \rpo s seem to be located
close to one another and their projections also qualitatively resemble
each other. These
are the \rpo s labelled with subscript \textit{F} in \reftab{t-data}.
We have
already noted that the periods of these \rpo s are approximately integer
multiples of the shortest one. Their qualitative similarities in
the \statesp\ projections of \reffig{f-PCAFunD} provide further
evidence that these orbits are related to one another, possibly
through sequences of bifurcations at other values of the \Reynolds\ number
\citep{ShWi16}.

\begin{figure}
	\centering
\begin{overpic}[width=0.45\textwidth]{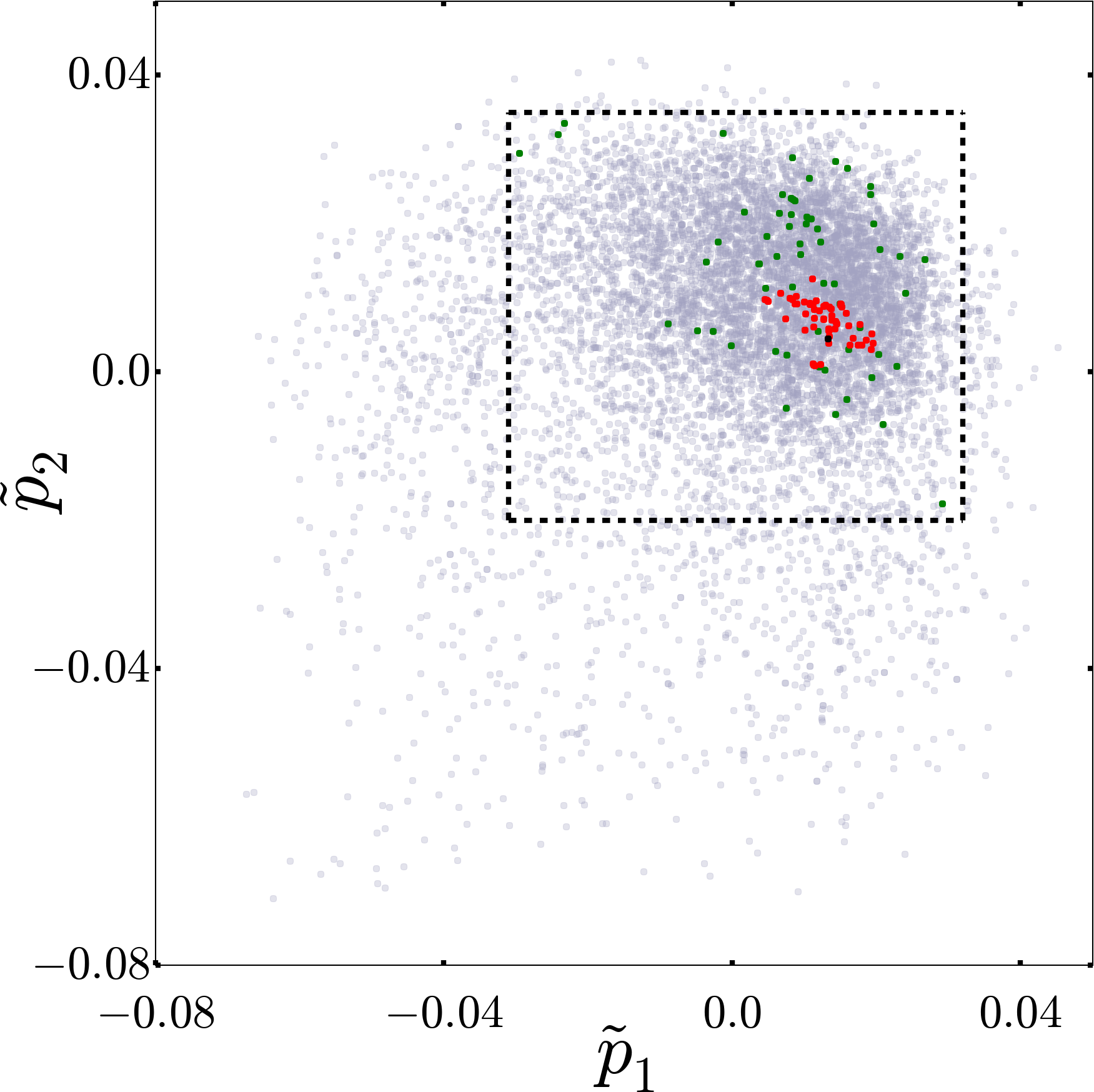}
	\put (0,0) {(a)}
\end{overpic}\quad
\begin{overpic}[width=0.45\textwidth]{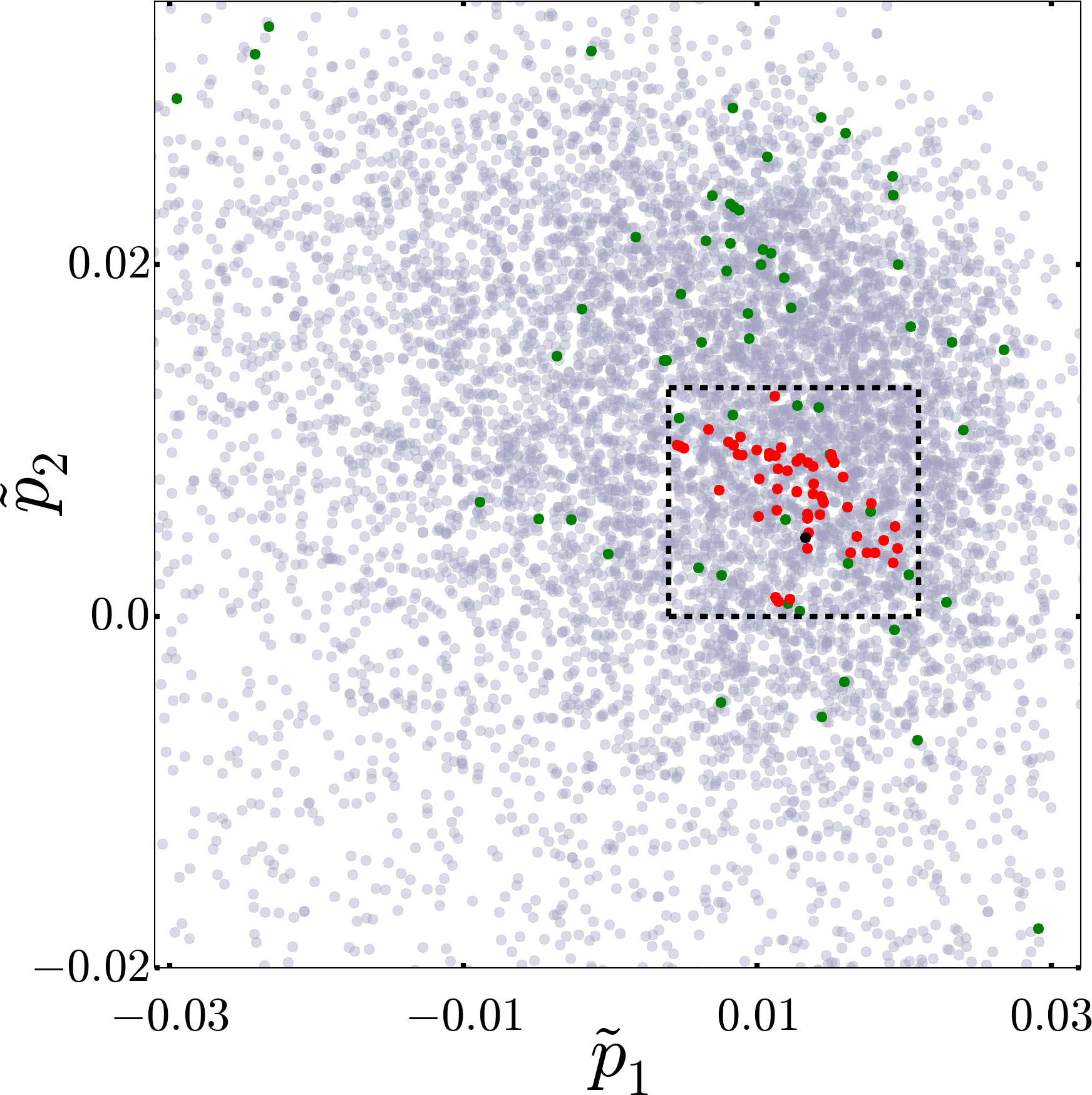}
	\put (0,0) {(b)}
\end{overpic}\\
\vspace{0.03\textwidth}
\begin{overpic}[width=0.45\textwidth]{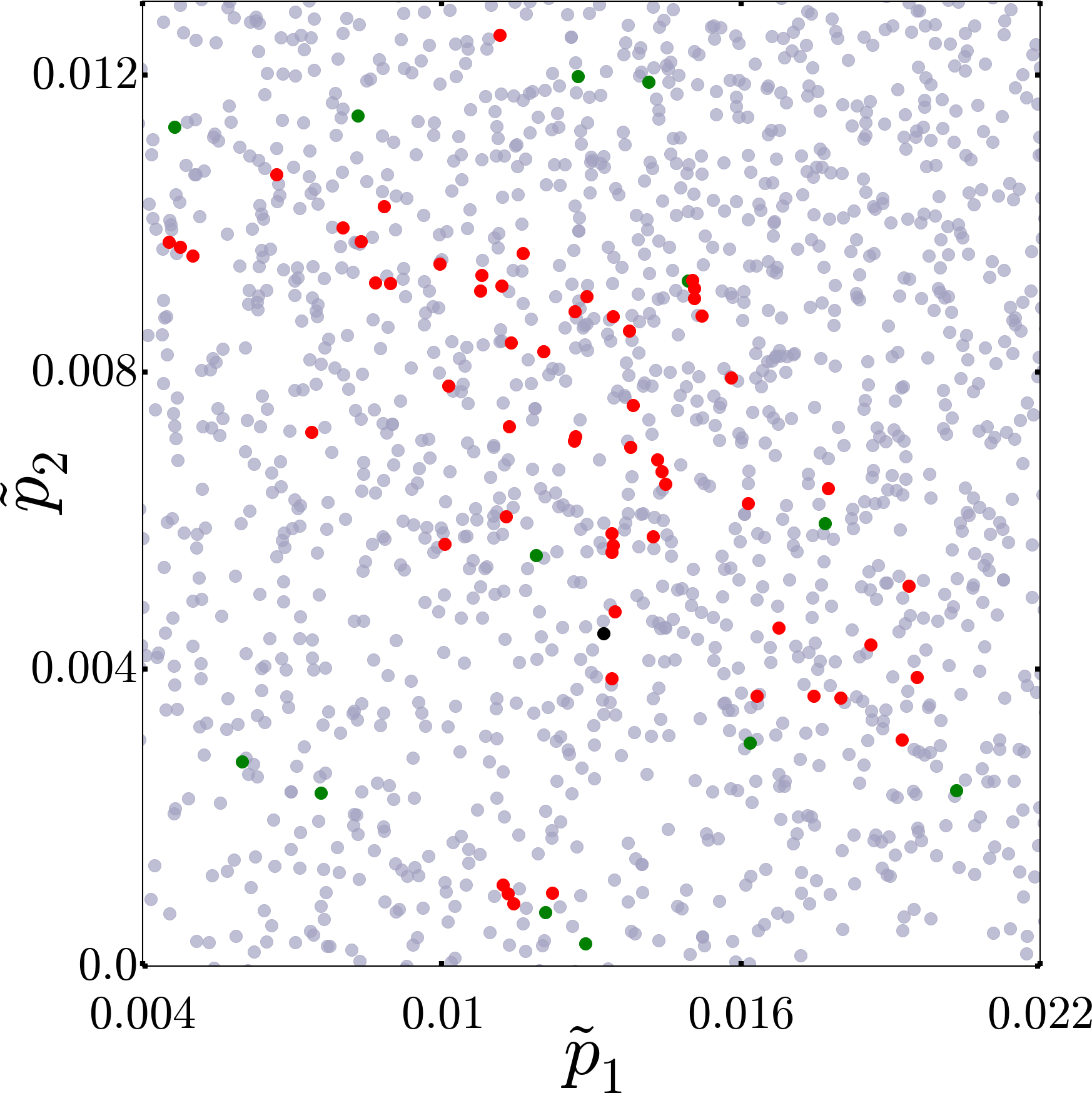}
	\put (0,0) {(c)}
\end{overpic}\quad
\begin{overpic}[width=0.45\textwidth]{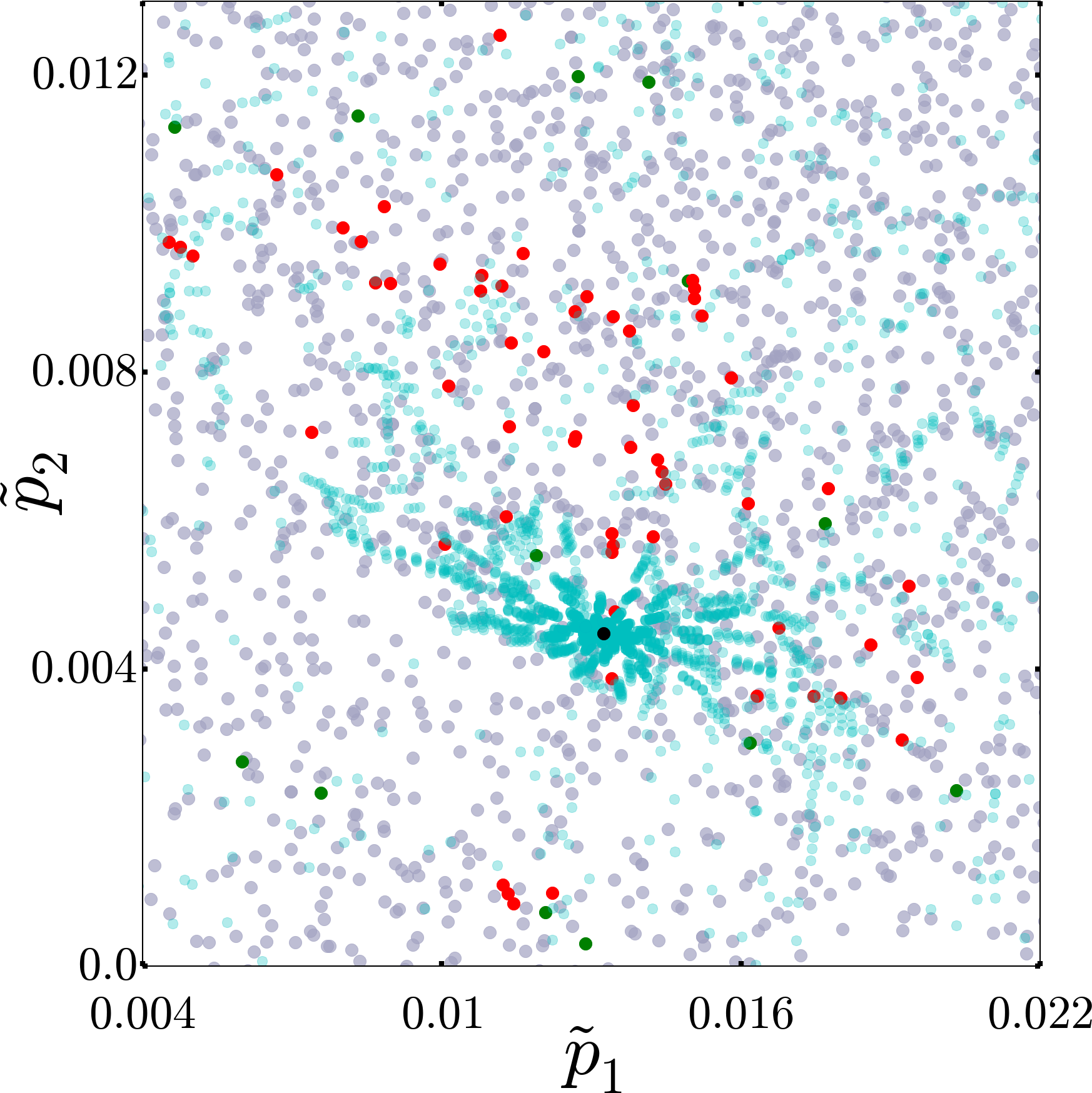}
	\put (0,0) {(d)}
\end{overpic}
	\caption[Global Poincar\'e section]{\label{f-Poincare}
    (Colour online)
(a) Turbulent trajectories (grey), and
\rpo s (black: \RPORUN{F/6.668}{6669},
		red: other members of the first family,
		green: the rest) on the {\PoincSec} \refeq{e-Psect}.
(b) Zoom-into the region enclosed by the dashed-rectangle on (a).
(c) Zoom-into the region enclosed by the dashed-rectangle on (b).
(d) Trajectories (cyan) on the unstable manifold of
\RPORUN{F/6.668}{6669} added to (c).
	}
\end{figure}

In order to develop more intuition about the \statesp\ geometry,
we reduce the flow further to a {\PoincSec} defined by
\beq
	\inprodE{\tilde{\ssp}_\PoincS - \expct{\tilde{\ssp}}}
            {\tilde{\vec{e}}'_3} = 0\, , \quad
	\inprodE{\velRed (\tilde{\ssp}_\PoincS)}{\tilde{\vec{e}}'_3} > 0
	\, . \label{e-Psect}
\eeq
In the projections of \reffig{f-PCAFunD}, the {\PoincSec}
\refeq{e-Psect} corresponds to $\tilde{p}_3 = 0$ plane, and for
visualisations of \reffig{f-Poincare}, we project intersections onto
the $(\tilde{\vec{e}}'_1, \tilde{\vec{e}}'_2)$ plane. However, it should
be noted that this {\PoincSec} is a codimension-1 hyperplane in
the symmetry reduced \statesp. \refFig{f-Poincare}\,({\it a}) shows $8560$
intersections (grey) of turbulent trajectories with the {\PoincSec}
\refeq{e-Psect} that were obtained from $147$ individual runs, along with
those (red and green) of \rpo s.

It is clear in \reffig{f-Poincare}\,({\it a}) that the turbulence visits the
region containing \rpo s more often than the rest of the \statesp . In
\reffig{f-Poincare}, the `first family' of \rpo s shown in \reffig{f-PCAFunD}
are marked red, except for \RPORUN{F/6.668}{6669}, the shortest one, which is
colored black.
\refFig{f-Poincare}\,({\it b}) is a close-up view of the region containing \rpo s,
enclosed by the dashed-rectangle in \reffig{f-Poincare}\,({\it a}) and
similarly, \reffig{f-Poincare}\,({\it c}) is a close-up of the region containing
the first family, marked with the dashed rectangle on
\reffig{f-Poincare}\,({\it b}). In \reffig{f-Poincare}\,({\it d}), we show orbits which
approximate the three-dimensional unstable manifold of
\RPORUN{F/6.668}{6669} overlaid over \reffig{f-Poincare}\,({\it c}).
(The computational aspects of tracing out the unstable manifold are
discussed in~\refsect{s-local}.)

The global visualisations of dynamics we presented above
provide us with insights about the \statesp\ structure: the most important
observation is that turbulent dynamics frequently visits the neighbourhoods of
the \rpo s found in this work. In addition, the ``first family''
\edit{orbits}---the set sharing
similar physical and stability properties---appear close to
each other in all projections of \reffig{f-PCAFunD} and
\reffig{f-Poincare}. Furthermore, the unstable manifold of the
shortest period member of this family visits the intersections of
the other members of the family on \reffig{f-PCAFunD}, which provides
additional evidence that these orbits emerge from a common
bifurcation sequence  \citep{ShWi16}.

Note that in the zoomed-in projection of the {\PoincSec},
\reffig{f-Poincare}\,({\it c}), some parts are often visited by the turbulent
trajectories (as indicated by the overlapping markers), while there are other
regions, even the ones that appear close to the frequently visited regions,
that tend to remain empty. This illustrates why measuring distances in the
\statesp\ of a turbulent fluid is a hard problem: Two points in \statesp\
that are seemingly very close to each other in $L_2$ or a similar norm may be
completely separated dynamically, since the geometry of the manifold on which
the turbulence takes place can be highly convoluted.

While the global visualisations give us a qualitative
view of the \statesp, they cannot tell us much about the finer
structure of the turbulent \statesp. This is not surprising since the
\statesp\ in question is very high dimensional; and it is
very unlikely that we can obtain a complete picture of dynamics from two-
and three-dimensional global visualisations. For a better understanding of the
\statesp\ geometry, one must study the neighbourhoods of important
\cohStr s individually, as illustrated in the next section.

\subsection{Local visualisations}
\label{s-local}

Global visualisations of the \statesp\ in \reffigs{f-PCAFunD}{f-Poincare}
support the earlier suggestion that the
members of the first family of \rpo s,
embedded in turbulence and listed separately in \reftab{t-data}, may be dynamically
related to each other.  The shortest period member \RPORUN{F/6.668}{6669} of first family
has three unstable ($|\Lambda_i| > 1$) Floquet multipliers:
\beq
    \Lambda_{1,2} = -0.1698 \pm i 1.418 \, , \quad
    \Lambda_{3} = -1.340 \, . \label{e-Floquet}
\eeq
This renders the associated unstable manifold of \RPORUN{F/6.668}{6669}
three-dimensional even after the symmetry reduction of the space and time
translation directions. Leading complex conjugate Floquet multipliers imply
spiral-out dynamics in the associated neighbourhood, while the negative-real third
Floquet multiplier implies that locally there exists a
topologically M\"{o}bius band-shaped dynamics such as the one observes in
period-doubling bifurcations. In the following, we numerically
approximate and visualize these one- and two-dimensional unstable sub-manifolds.

\rf{BudCvi15} numerically approximated the one- and
two-dimensional unstable manifolds of \rpo s in a \KS\ system. In those
computations, a local {\PoincSec} was constructed in the neighbourhood
of a \po\ where they initiated orbits whose dynamics approximately
covered the linear unstable manifold; hence their forward integration
approximated the unstable manifold away from the linearized neighbourhood.
This strategy was adapted for calculating one-dimensional unstable manifold of
the localized ``edge state'' \rpo\ of the pipe flow \edit{in \refref{BudHof17}, in order}. 
We apply here this method to a \rpo\
embedded in turbulence, with a three-dimensional unstable manifold, a case
that was not considered in the aforementioned studies.

To this end, we first define a local {\PoincSec} in the neighbourhood of
\RPORUN{F/6.668}{6669} as the half-hyperplane
\beq
	\inprodE{\sspRed_\PoincS - \sspRed_p}
            {\velRed (\sspRed_p)}           = 0\, , \quad
	\inprodE{\velRed (\sspRed_\PoincS)}
            {\velRed (\sspRed_p)}           > 0
	\, , \label{e-PsectLocal}
\eeq
where $\sspRed_p$ is a point on \RPORUN{F/6.668}{6669},
which we have
arbitrarily chosen as its intersection with the global {\PoincSec}
\refeq{e-Psect}. The \rpo\ \RPORUN{F/6.668}{6669} is a fixed point
of the Poincar\'e map on the section \refeq{e-PsectLocal} with stability
multipliers equal to its Floquet multipliers. Associated Floquet vectors,
however, need to be projected onto this section. This is a two-stage process
since we compute the Floquet vectors as eigenfunctions
$\FloquetV_i$ of the eigenvalue problem
\beq
 \LieEl_z (- l_{p}) \frac{d \flow{\period{p}}{\vec{\hat{u}}_p}}{
                            d \vec{\hat{u}}_p}
 \FloquetV_i = \Lambda_i \FloquetV_i \, ,
\eeq
in full \statesp . First we project these vectors onto the slice as
\beq
	\FloquetVRed_{i} = \FloquetV_{i}
- \frac{
	\inprodE{\sliceTan{}}{\FloquetV_{i}}
}{
	\inprodE{\groupTan (\sspRed_{p})}{\sliceTan{}}
 } \groupTan (\sspRed_{p}) ,
\label{e-projSlice}
\eeq
where $\FloquetVRed_{i}$ denotes the projected vector on the slice.
Then we project translation-symmetry reduced Floquet vectors onto the
{\PoincSec} as
\beq
	\FloquetVRed_{i, \PoincS}
	= \FloquetVRed_{i}
	-\frac{
	 \inprodE{\velRed (\sspRed_p)}{
			   \FloquetVRed_{i}}
	       }{
	       \inprodE{\velRed (\sspRed_p)}{
	       	\velRed (\sspRed_p)}
	       } \velRed (\sspRed_p) \, .
\label{e-projPoincare}
\eeq
Projections \refeq{e-projSlice} and \refeq{e-projPoincare} onto slice
hyperplane \refeq{eq:slicecond} and onto the {\PoincSec} hyperplane
\refeq{e-PsectLocal} follow the same geometrical principle as the
Appendix of \citep{BudHof17}.

\begin{figure}
	\centering
\begin{overpic}[width=0.45\textwidth]{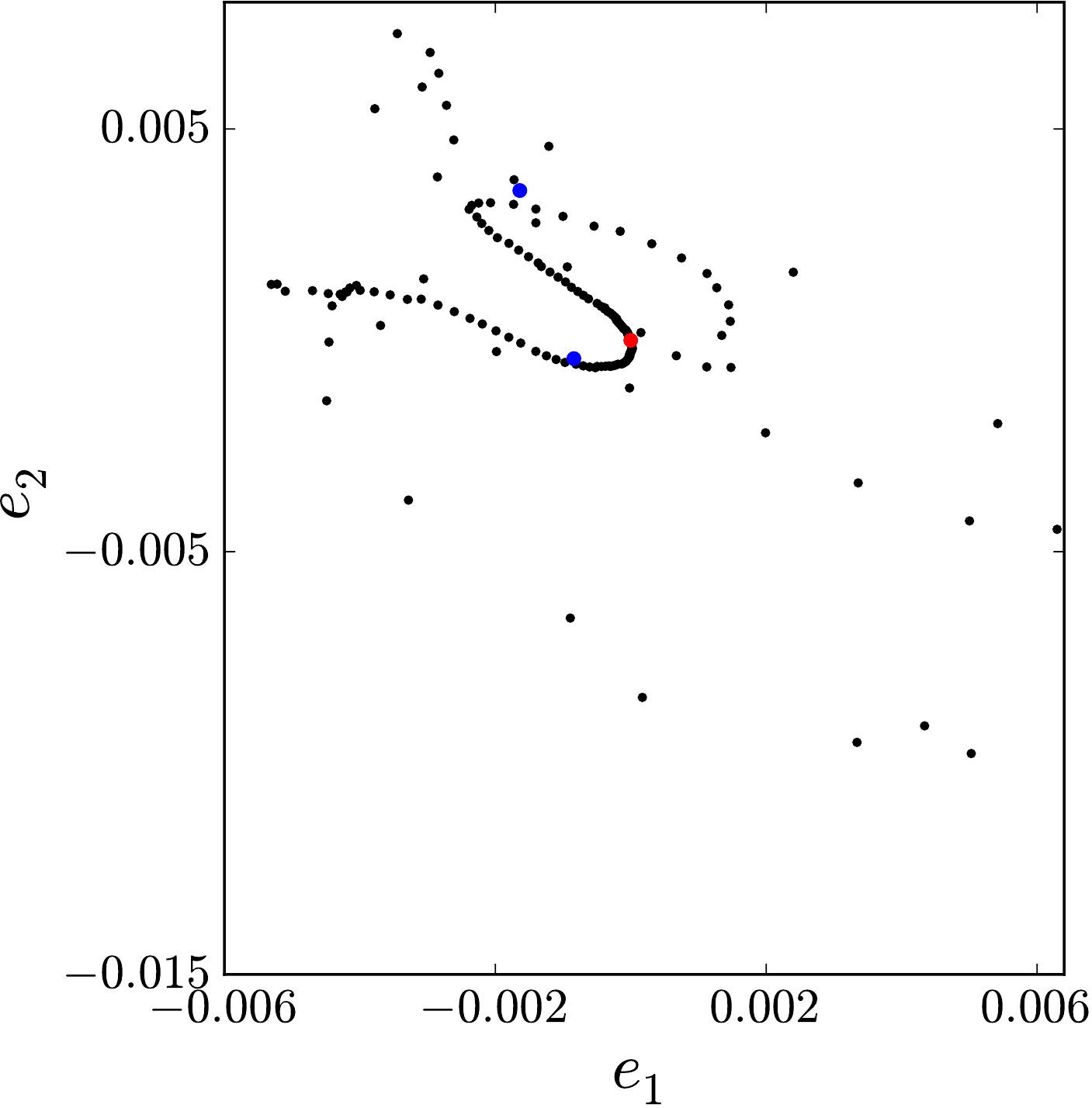}
	\put (0,0) {(a)}
\end{overpic}\quad
\begin{overpic}[width=0.45\textwidth]{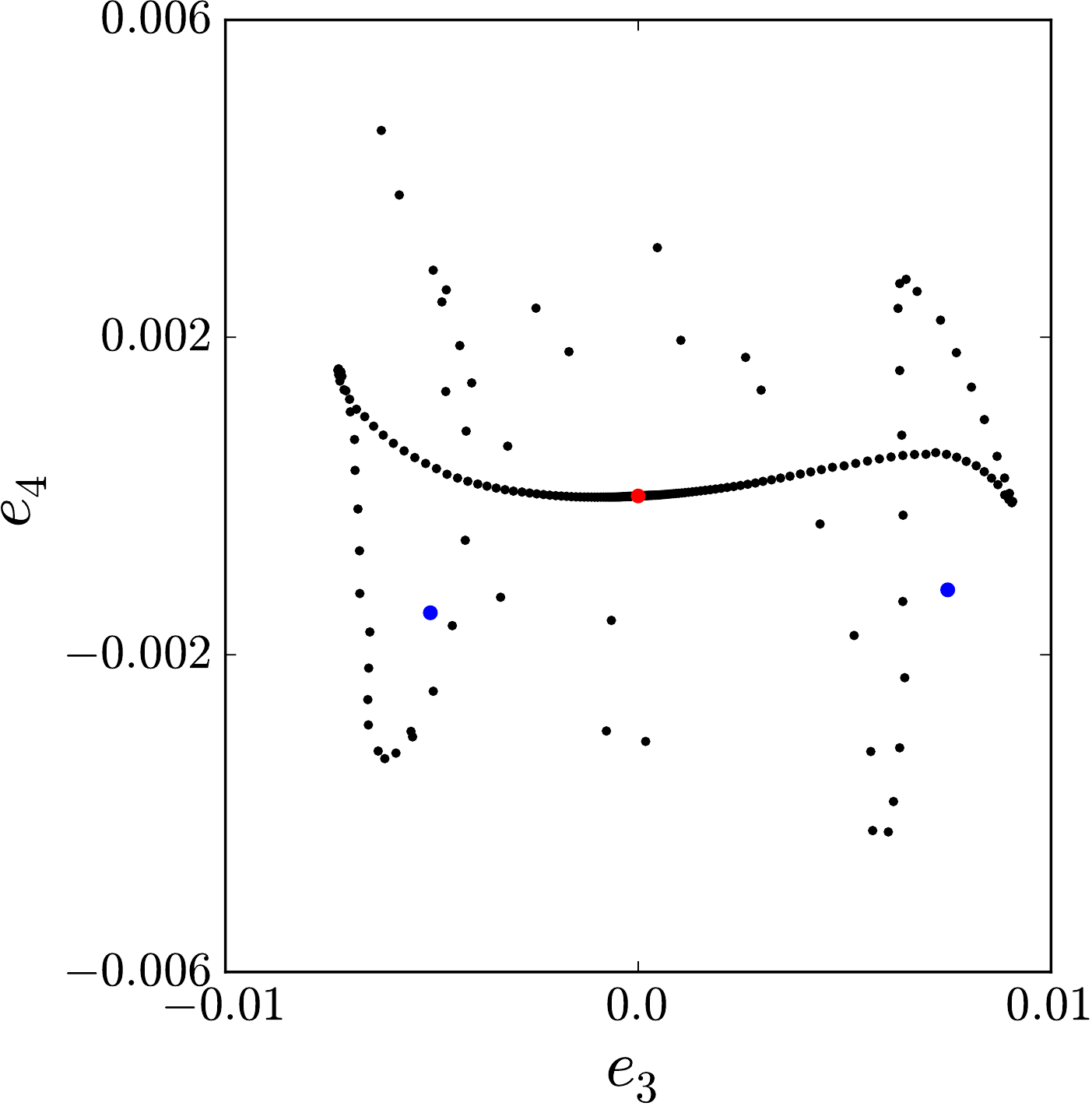}
	\put (0,0) {(b)}
\end{overpic}\\
\vspace{0.03\textwidth}
\begin{overpic}[width=0.45\textwidth]{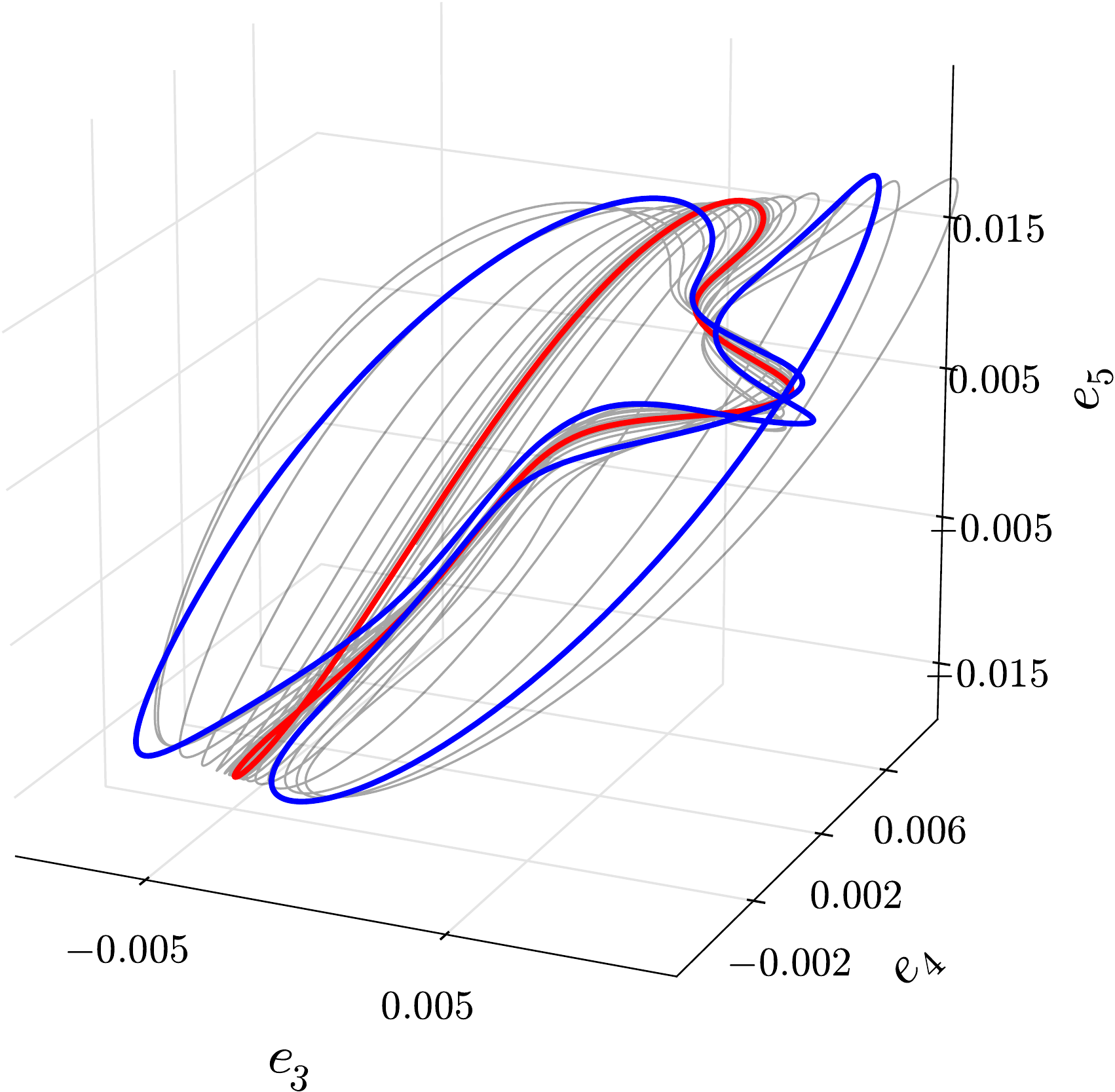}
	\put (0,0) {(c)}
\end{overpic}\quad
\begin{overpic}[width=0.45\textwidth]{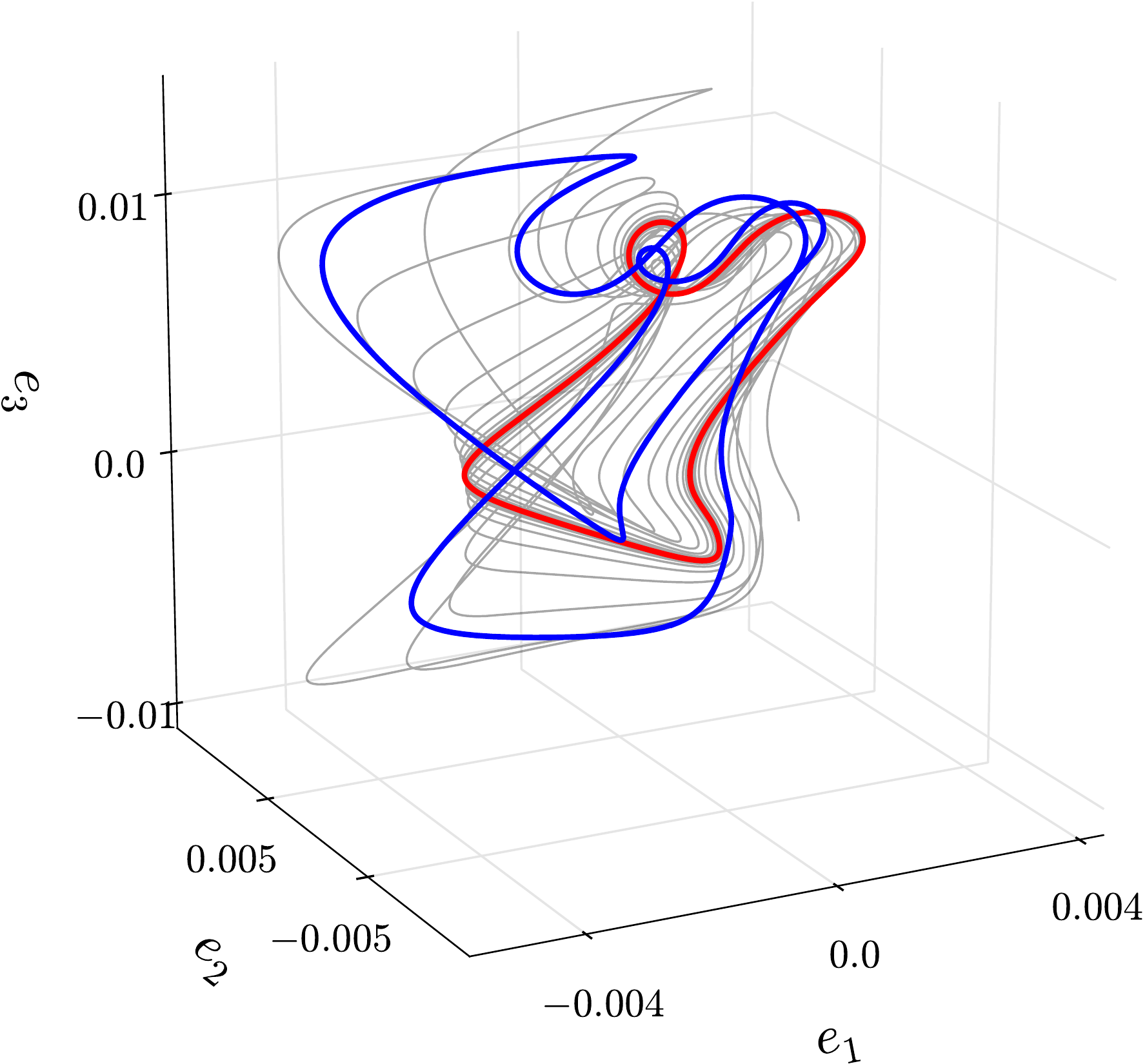}
	\put (0,0) {(d)}
\end{overpic}
	\caption[Global Poincar\'e section]{\label{f-PoincareLocal}
    (Colour online)
		(a,b)
One-dimensional submanifold in the unstable manifold of
\RPORUN{F/6.668}{6669} capturing locally linear dynamics in the
$\FloquetVRed_{3, \PoincS}$ direction in the {\PoincSec}, two
projections onto local coordinates \refeq{e-ProjBases}.
Fixed point corresponding to \RPORUN{F/6.668}{6669} is at the origin (red),
and the 2-cycle \RPORUN{F/13.195}{6652} is marked blue.
(c,d)~Three-dimensional projections of one ($\delta = 0$) of the $8$
trajectories in panels~({\it a,b}).
	}
\end{figure}

In order to visualise the one-dimensional unstable submanifold of
\RPORUN{F/6.668}{6669}, we initiate trajectories
from the points
\beq
\sspRed_\PoincS (\delta)
= \sspRed_{p} \pm \epsilon
 |\Lambda_3|^{\delta} \FloquetVRed_{3, \PoincS}
\, , \quad \mbox{where} \quad \delta \in [0,1) \,.
\label{e-UnstMan3}
\eeq
These initial conditions approximately cover the locally linear
one-dimensional piece of the unstable manifold in
$\FloquetVRed_{3, \PoincS}$ direction such that first-return of
$\tilde{\ssp}_\PoincS (0)$ coincides with initial location of
$\tilde{\ssp}_\PoincS (1)$. We discretised \refeq{e-UnstMan3}
by choosing four equidistant points in $\delta$ and
set $\epsilon = 10^{-3}$
(Floquet vectors are normalized such that
$\normE{\ssp_p} = \normE{\FloquetV_i}$). We
forward-integrate these initial conditions while recording their
intersections with the {\PoincSec} \refeq{e-PsectLocal}. 	
\refFig{f-PoincareLocal}\,({\it a,b}) shows the first $25$ intersections
of these orbits with the {\PoincSec} on two-dimensional
projections and panels ({\it c,d}) shows one of these orbits in a
three-dimensional projection along with
\RPORUN{F/6.668}{6669} and \RPORUN{F/13.195}{6652}.
The origin of the projections in \reffig{f-PoincareLocal}
is $\sspRed_{p}$ and the projection coordinates
are
\bea
e_1 &=& \inprodE{\sspRed}{\Re \FloquetVRed_{1,\bot}}
\,,\quad
e_2 \,=\, \inprodE{\sspRed}{\Im \FloquetVRed_{1,\bot}},
\continue
e_3 &=& \inprodE{\sspRed}{\FloquetVRed_{3,\bot}}
\,,\qquad
e_4 \,=\, \inprodE{\sspRed}{\FloquetVRed_{6,\bot}},
 \label{e-ProjBases}\\
e_5 &=& \inprodE{\sspRed}{
	\velRed (\sspRed_p) /
	\normE{\velRed (\sspRed_p)}},
\nonumber
\eea
where $\FloquetVRed_{6,\bot}$ is the symmetry-reduced Floquet vector
in the least stable direction (comes after marginal axial and temporal
translation directions), and subscript $\bot$ indicates that these
vectors are Gram-Schmidt orthonormalized.

\refFig{f-PoincareLocal}\,({\it a,b}) shows that one-dimensional shape of locally
linear dynamics is preserved as it is extended far away from the origin.
Trajectories in \reffig{f-PoincareLocal}\,({\it a,b}) spread in a
higher-dimensional manifold once they reach the neighbourhood of
period-doubled \RPORUN{F/13.195}{6652}. For additional comparison,
in \reffig{f-PoincareLocal}\,({\it c,d}) we plot
different three-dimensional projections of \RPORUN{F/6.668}{6669}, the
$\delta = 0$ in \refeq{e-UnstMan3} orbit, and \RPORUN{F/13.195}{6652} in different
three-dimensional projections. Qualitative
similarities between the shape of the unstable manifold and
\RPORUN{F/13.195}{6652} are remarkable. For a quantitative conclusion,
one should search for a heteroclinic connection from three-dimensional
unstable manifold of \RPORUN{F/6.668}{6669} to the \rpo\
\RPORUN{F/13.195}{6652}. That, however, is beyond the scope of the current
work.

\begin{figure}
	\centering
\begin{overpic}[width=0.3\textwidth]{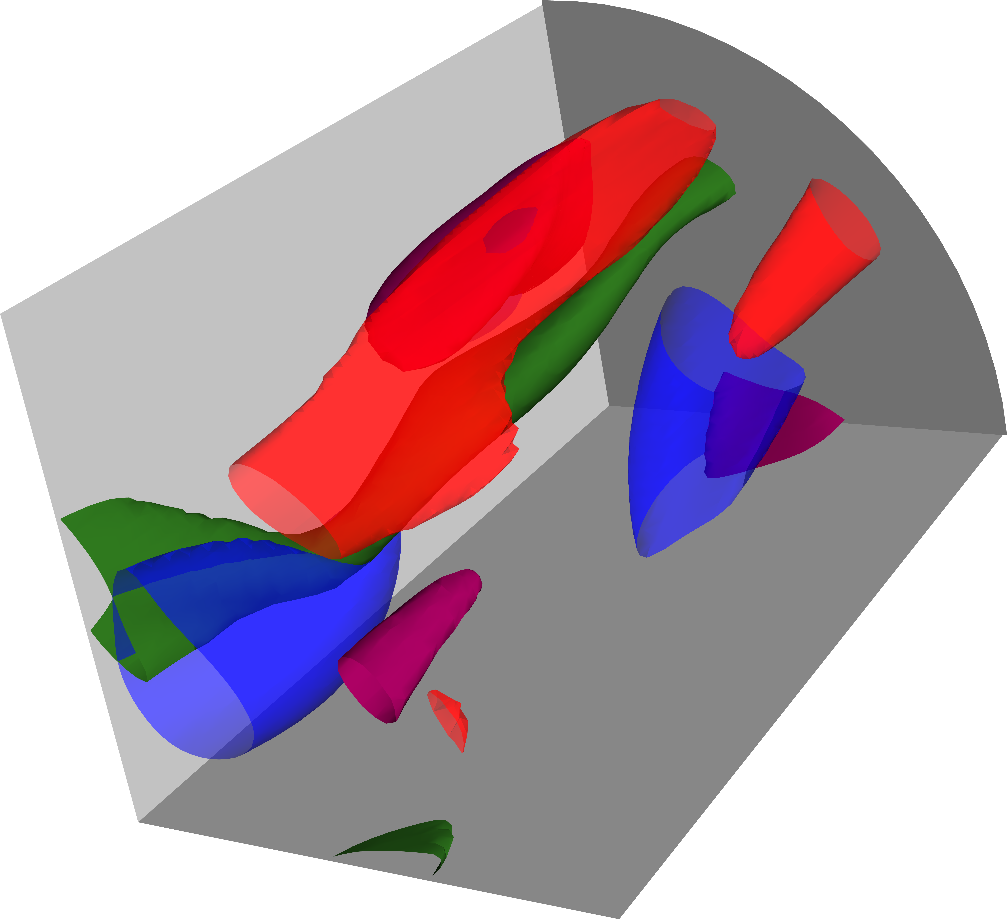}
	\put (0,0) {(a)}
\end{overpic}\quad
\begin{overpic}[width=0.3\textwidth]{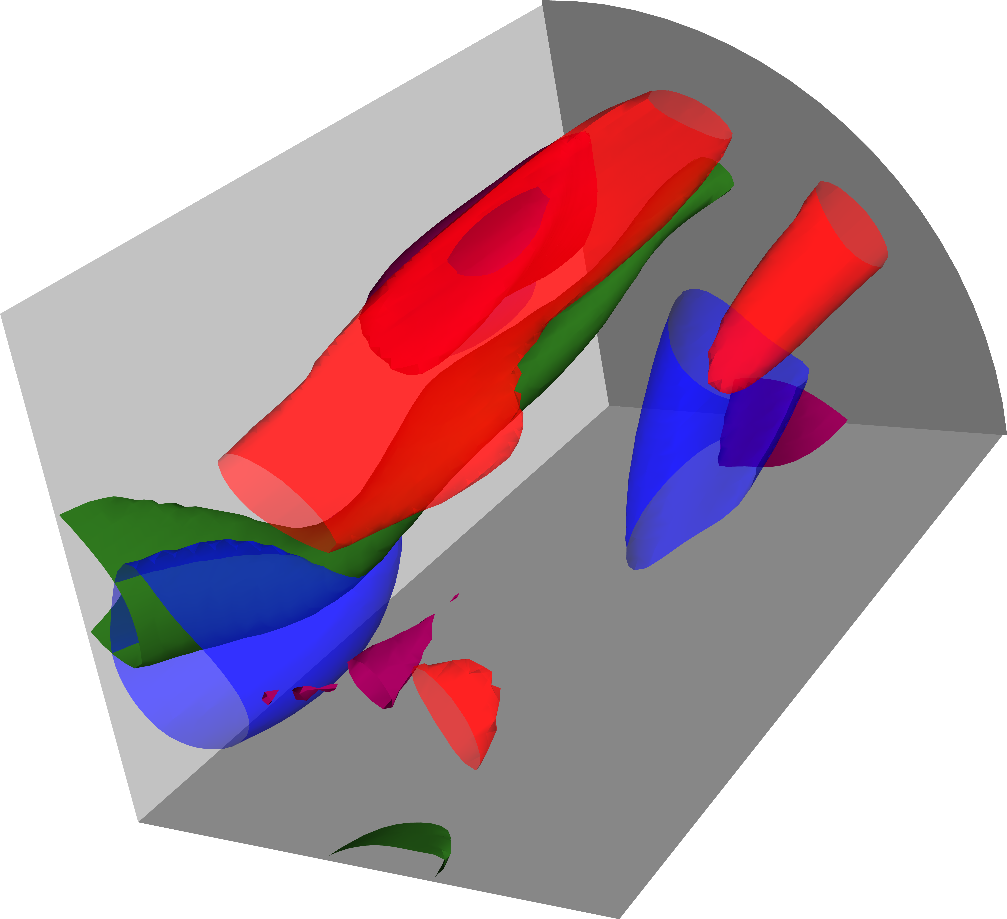}
	\put (0,0) {(b)}
\end{overpic}
\begin{overpic}[width=0.3\textwidth]{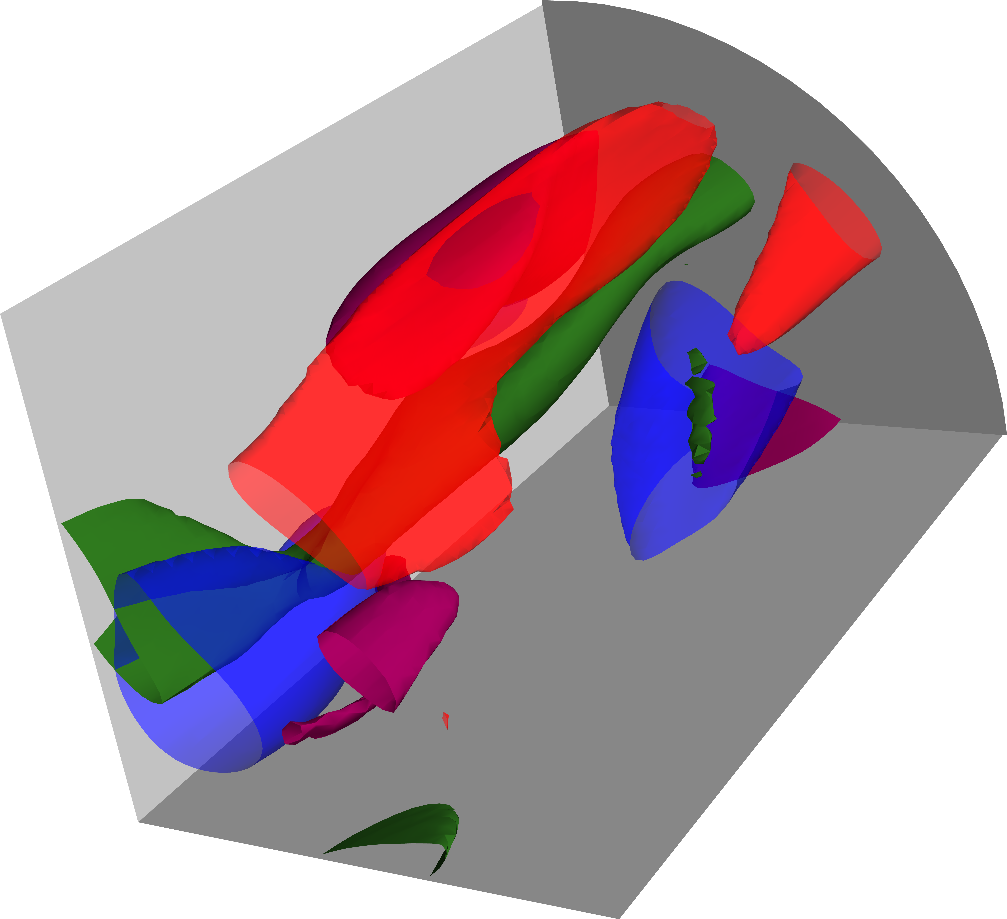}
	\put (0,0) {(c)}
\end{overpic}
\\
\vspace{0.03\textwidth}
\begin{overpic}[width=0.3\textwidth]{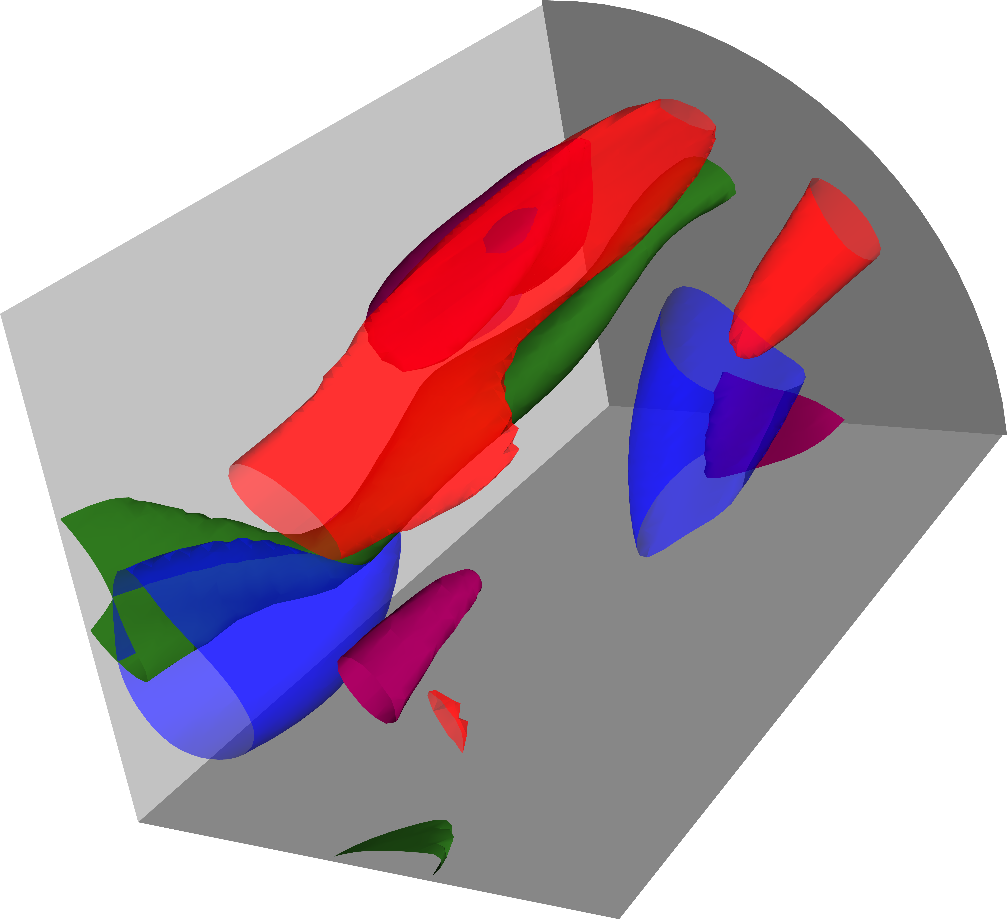}
	\put (0,0) {(d)}
\end{overpic}\quad
\begin{overpic}[width=0.3\textwidth]{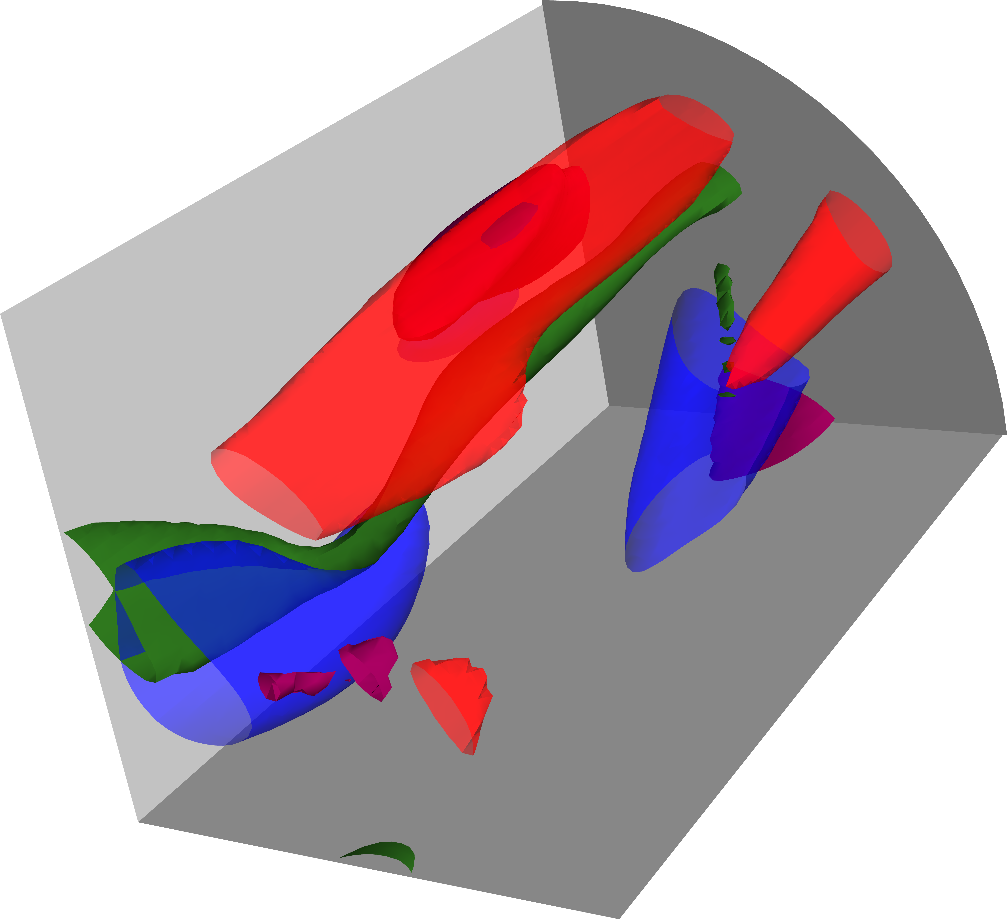}
	\put (0,0) {(e)}
\end{overpic}
\begin{overpic}[width=0.3\textwidth]{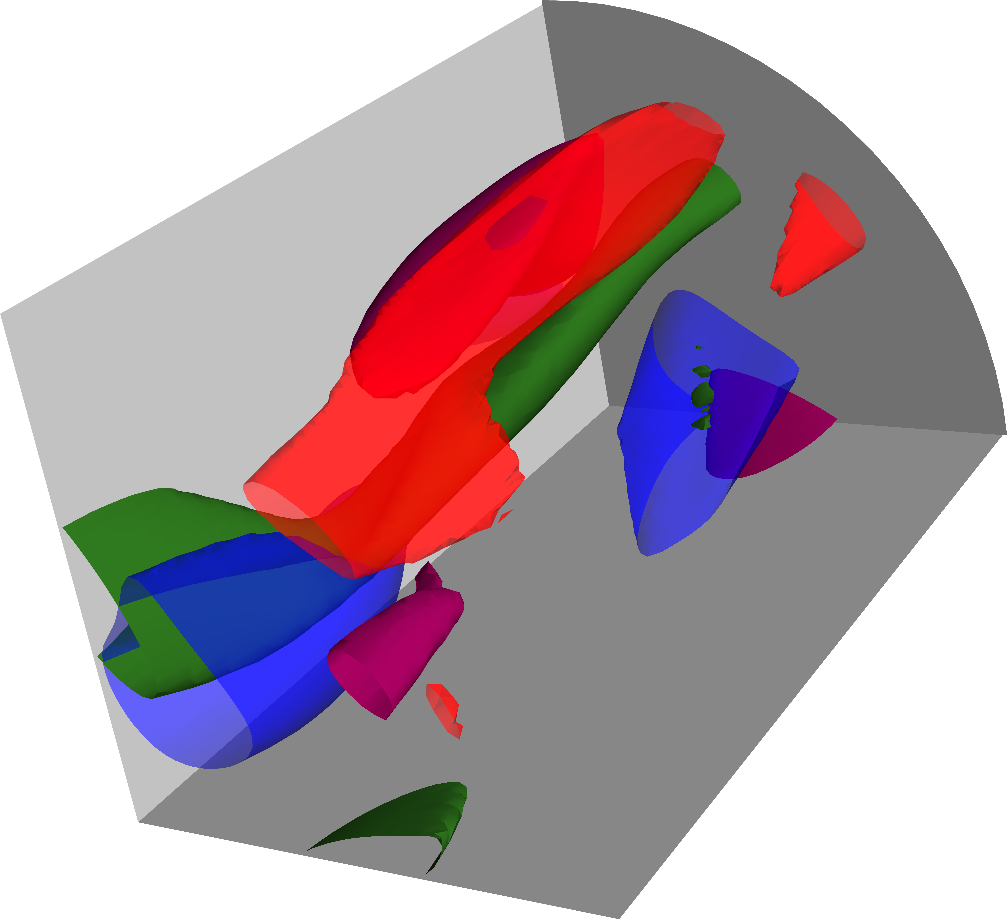}
	\put (0,0) {(f)}
\end{overpic}
	\caption[Flow structures]{\label{f-PoincareStruct}
    (Colour online)
        Streamwise velocity isosurfaces at $u = \pm 0.5U_b$ (red and blue)
        and streamwise vorticity isosurfaces at $\omega_z = \pm 2.4 (U_b/D)$
        (purple and green) of
		(a) \RPORUN{F/6.668}{6669};
		(b,c) \RPORUN{F/13.195}{6652}; and
        (d,e,f)
        the unstable manifold of \RPORUN{F/6.668}{6669} at discrete times
        $n = 0, 23, 24$.
	}
\end{figure}

For further comparison, we visualise the streamwise velocity and
vorticity isosurfaces of
\RPORUN{F/6.668}{6669}, \RPORUN{F/13.195}{6652}, and three-snapshots
on the unstable manifold of \RPORUN{F/6.668}{6669} in
\reffig{f-PoincareStruct}.
All panels of \reffig{f-PoincareStruct}
correspond to their respective intersections with the {\PoincSec}
\refeq{e-PsectLocal}, and only one-eighth
(one-quarter in azimuthal and one-half in axial directions)
of the pipe is shown. The one-eighth visualisation suffices, since
we work in the subspace
with 4-fold symmetry in the azimuthal direction, and the other half
of the pipe in the axial direction can be obtained from the first half by
the shift-and-reflect \refeq{e-ShiftNRef} symmetry.

While all panels of \reffig{f-PoincareStruct} appear
similar, they differ in details. \refFig{f-PoincareStruct}\,({\it d}),
the initial point on the unstable manifold, is virtually
indistinguishable from \RPORUN{F/6.668}{6669} in \reffig{f-PoincareStruct}\,({\it a}).
Flow structures of
\RPORUN{F/13.195}{6652} at its two-intersections  with the
{\PoincSec}, \reffig{f-PoincareStruct}\,({\it b,c}),
differ from \RPORUN{F/6.668}{6669} and from each-other
only in minute details; one has to compare the streak and roll sizes one-by-one.
These nuances are reflected on the selected points on the unstable
manifold shown in \reffig{f-PoincareStruct}\,({\it d,e,f}),
although only identifiable after a careful
inspection. These difficulties illustrate the power of \statesp\
visualisation (cf. \reffig{f-PoincareLocal}), without which the relation of
\RPORUN{F/6.668}{6669} to \RPORUN{F/13.195}{6652} would have been very
hard to elucidate.

\begin{figure}
	\centering
\begin{overpic}[width=0.45\textwidth]{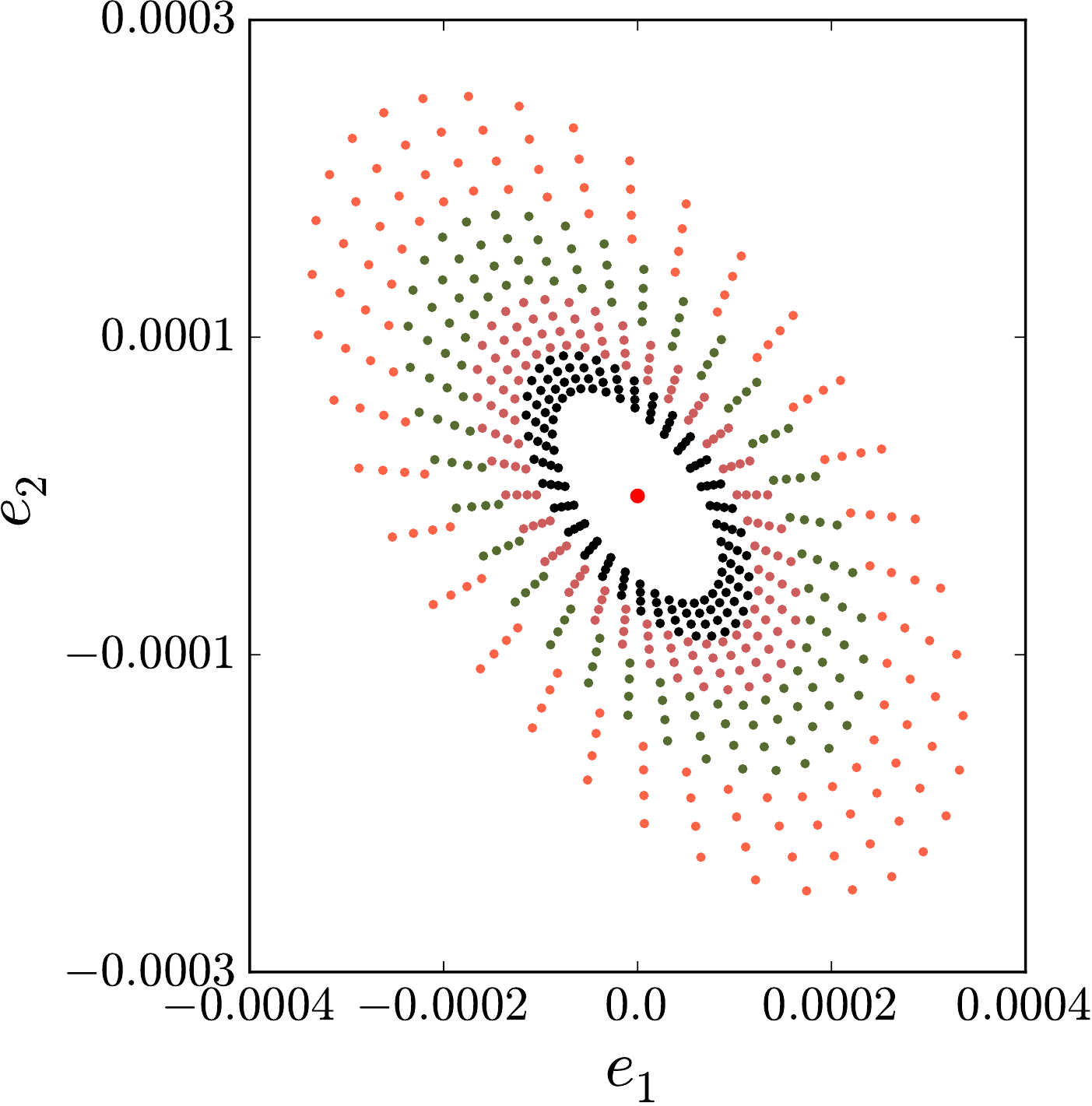}
	\put (0,0) {(a)}
\end{overpic}\quad
\begin{overpic}[width=0.45\textwidth]{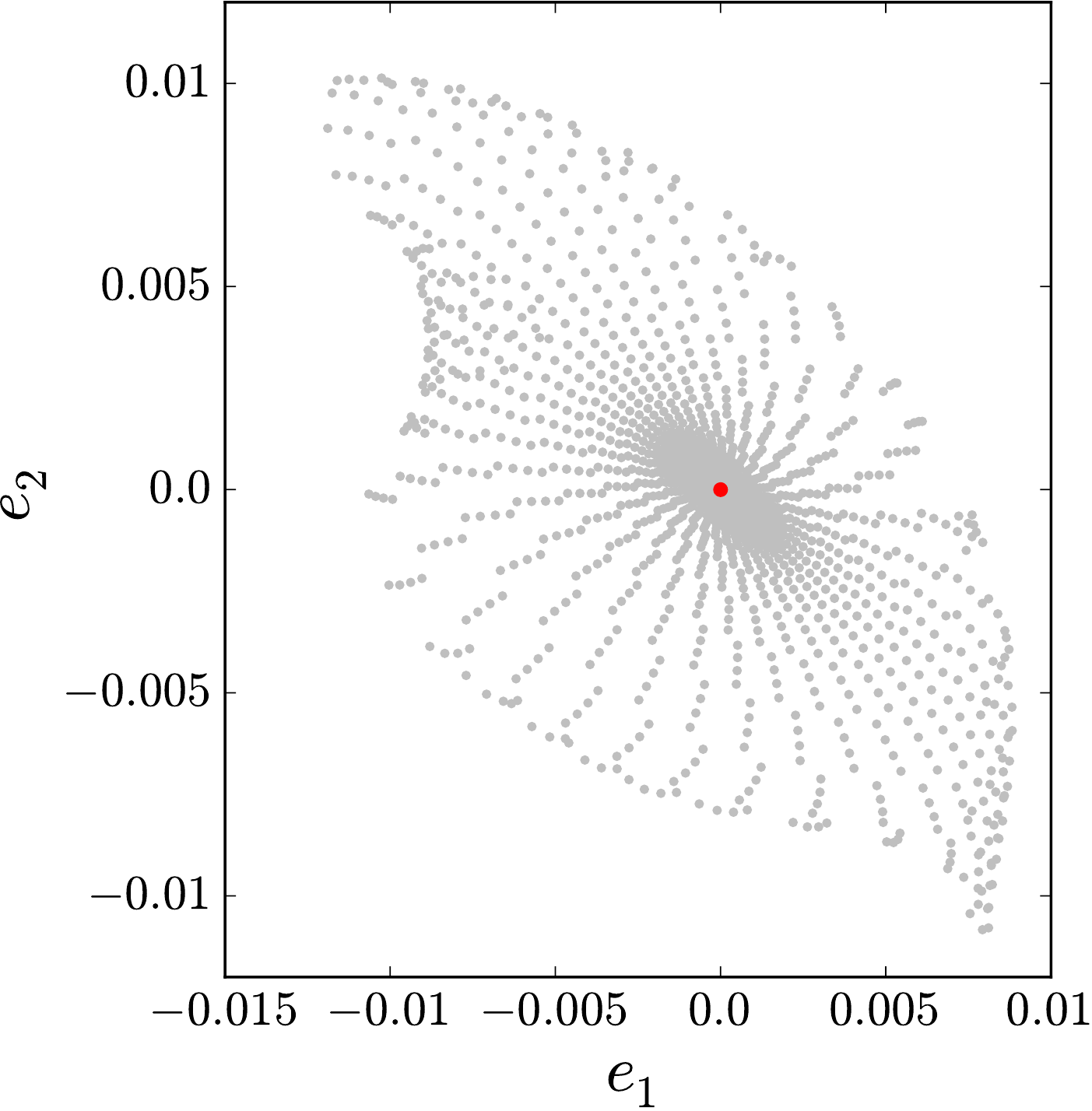}
	\put (0,0) {(b)}
\end{overpic}
	\caption[2D submanifold]{\label{f-UnstMan2D}
    (Colour online)
Two-dimensional submanifold in the unstable manifold of
\RPORUN{F/6.668}{6669} capturing locally linear dynamics in the
$\Re \FloquetVRed_{1, \PoincS}, \Im \FloquetVRed_{1, \PoincS}$ plane
in the {\PoincSec} projected onto local coordinates
\refeq{e-ProjBases}. Panel (a) shows initial conditions (black) and
their orbits' first $3$ intersections with the {\PoincSec},
in different colours. $15$ intersections shown in (b) illustrating
global shape of the $2D$ submanifold.
	}
\end{figure}

A set of initial conditions that approximately covers the linearised
dynamics in the plane
$(\Re \FloquetVRed_{1, \PoincS}, \Im \FloquetVRed_{1, \PoincS})$
is given by
\beq
	\tilde{\ssp}_\PoincS (\phi, \delta)
	= \hat{\ssp}_{p} \pm \epsilon
	|\Lambda_1|^{\delta}
	(
	\Re \FloquetVRed_{1, \PoincS} \cos \phi
  + \Im \FloquetVRed_{1, \PoincS} \sin \phi
	)
	\, , \quad
	\delta \in [0,1) \, ,\quad \phi \in [0, 2 \pi).
	\label{e-UnstMan2D}	
\eeq
We discretise \refeq{e-UnstMan2D} by choosing $4$
equidistant points in $\delta$ and $36$ points in $\phi$
and set $\epsilon = 10^{-3}$. First three intersection
of these initial conditions with the {\PoincSec}
\refeq{e-PsectLocal} are visualised in the projection
\reffig{f-UnstMan2D}\,({\it a}) in different colours,
where black points correspond to the initial
conditions. This figure illustrates the motivation for the
particular approximation: Initial conditions
\refeq{e-UnstMan2D} define an elliptic band in the
$(\Re \FloquetVRed_{1, \PoincS}, \Im \FloquetVRed_{1, \PoincS})$
plane, such that the inner ellipse is mapped to the outer one
by the linearised dynamics on the {\PoincSec}. The
totality of these initial conditions captures well the linearised
dynamics in this neighbourhood.

\refFig{f-UnstMan2D}\,({\it a}) also
illustrates the validity of linearised dynamics as each initial
condition simply expands and rotates according to real and imaginary
part of $\FloquetV_1$, when their distance to $\sspRed_p$ is of
order $10^{-4}$. In \reffig{f-UnstMan2D}\,({\it b}), we show the same
projection for $15$ intersections of these orbits on the {\PoincSec}
as they leave the neighbourhood of the \rpo. At this stage,
the shape is no longer an ellipse but it is starting to develop
corners, possibly due to being distorted by a stable manifold.
It should be noted, however, that the sub-manifold associated with the
linearised dynamics on the plane
$(\Re \FloquetVRed_{1, \PoincS}, \Im \FloquetVRed_{1, \PoincS})$
is still two-dimensional. Note that the scales of axes in
\reffig{f-UnstMan2D}\,({\it b}) are about two orders of magnitude larger than
those on \reffig{f-UnstMan2D}\,({\it a}), and also  that they are comparable to
scales of \reffig{f-Poincare}\,({\it d}). In other words, a \rpo\ not only
guides the dynamics in its immediate neighbourhood, but it indeed guides,
through its unstable manifold, nearby motions at considerable finite distances.

\begin{figure}
 \centering
   (a) \includegraphics[height=0.45\textwidth]{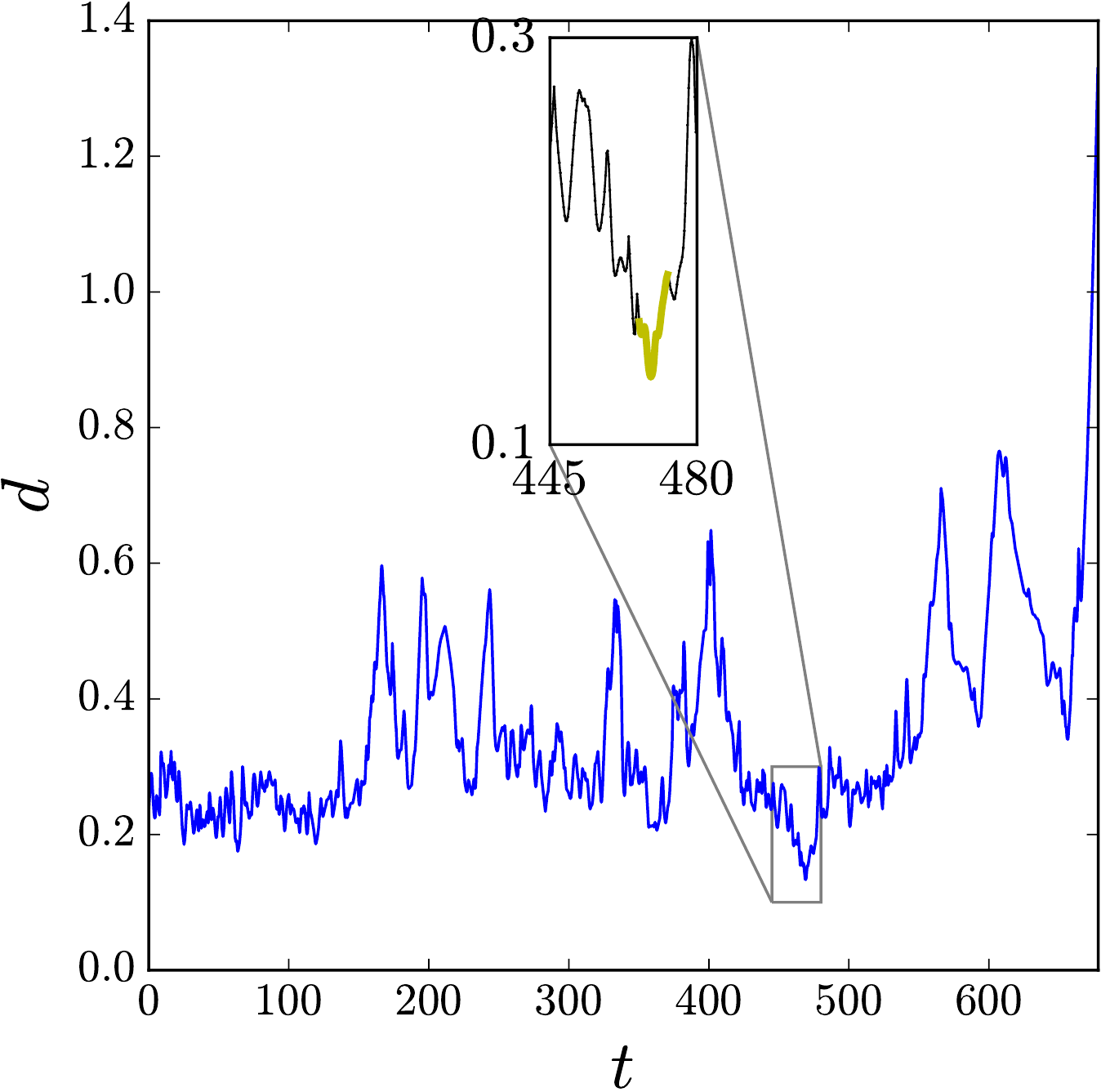}
  ~(b) \includegraphics[height=0.45\textwidth]{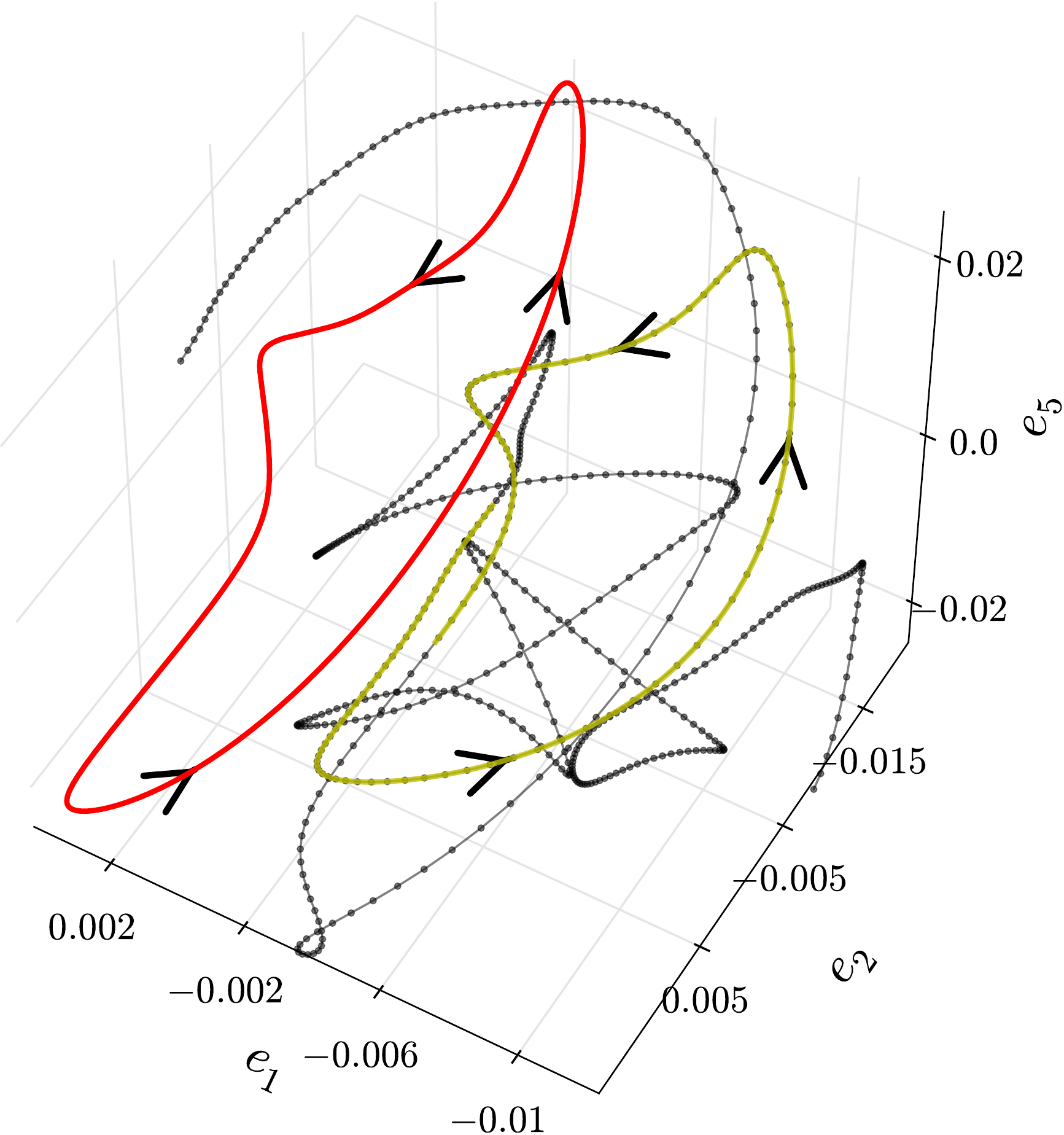}
\caption{\label{f-projMinDist}
\edit{
        (Colour online)
(a)
Minimum distance \refeq{eqProjMinDist} between a long turbulent
trajectory $\sspRed_{turb}(t')$ and the \rpo\ \RPORUN{F/6.668}{6669}.
Inset: A time interval containing the closest approach.
(b)
\RPORUN{F/6.668}{6669} (red) and a shadowing segment (corresponding to
the inset of panel a) of a turbulent trajectory (dotted line) visualized
as a projection onto \refeq{e-ProjBases}. The closest approach of the
turbulent trajectory to \RPORUN{F/6.668}{6669}, $d < 0.15$, is
highlighted yellow.
     }
        }
\end{figure}

\edit{
The above \PoincSec s illustrate the ways in which a \rpo\ shapes the
geometry of its immediate neighborhood. However, as in the example at
hand the unstable manifold \PoincSec\ is three\dmn, it is hard to discern
any structure in the ergodic sea in two\dmn\ projections such as
\reffig{f-Poincare}: in all our \PoincSec s the ergodic sea appears to be
structureless cloud, exhibiting no foliation typical of
--let's say-- Lorenz attractor or \KS\ attractor \citep{BudanurThesis}.
The influence of a \rpo\ is here easier to visualize by studying shadowing
episodes, \ie, the turbulent trajectory's visits to a given \rpo's
neighbourhood, such as \reffig{f-projMinDist}\,(a).
Here we have defined the minimum distance between trajectories labelled
`$turb$' and `$\textrm{RPO}$' in the fully symmetry-reduced \statesp\
(continuous symmetry reduced by \refeq{e-fFsliceShift}, the discrete
half-rotation symmetry $\LieEl_{\theta}$ \refeq{e-GSRpipe} reduced to the
fundamental domain), measured in the energy norm \refeq{e-normE}, as
\beq
d(\zeit)= \min_{\zeit'\in[0,\period{\textrm{RPO}}]}
       \frac{\normE{\tilde{\ssp}_{turb}(\zeit)-\tilde{\ssp}_{\textrm{RPO}}(\zeit')}}
            {\normE{\tilde{\ssp}_{turb}(\zeit)}}
\,.
\ee{eqProjMinDist}
Compared to the typical \RPORUN{F/6.668}{6669} linearized neighborhood
scales (see \reffig{f-PoincareLocal}), the yellow shadow in
\reffig{f-projMinDist}\,(b) is a considerable distance away, but it still
completes one co-rotating shadowing period very nicely.
Such shadowing
episodes offer further support to our main thesis, that \rpo s, together
with their stable/unstable manifolds, shape the \stateDsp\ dynamics
within their local neighborhoods.
    }

\subsection{Local visualisation: energy norm vs. \lowpass\ norm}
\label{s:TW-norm}

As our final example of a local visualization of the \statesp, we
examine the local unstable manifold of a \reqv.
The primary goal here is to show with this example is how profoundly
the choice of the inner product (or norm) can affect
the visualisation, and the conclusions drawn from it.
Since in the slice the \reqva\ reduce to \eqva,
there is no need for a {\PoincSec}.
Therefore, visualizing low-dimensional unstable manifolds of \reqva\
is more straightforward compared to those of \rpo s discussed above.

\begin{figure}
	(a)\includegraphics[width=.48\textwidth]{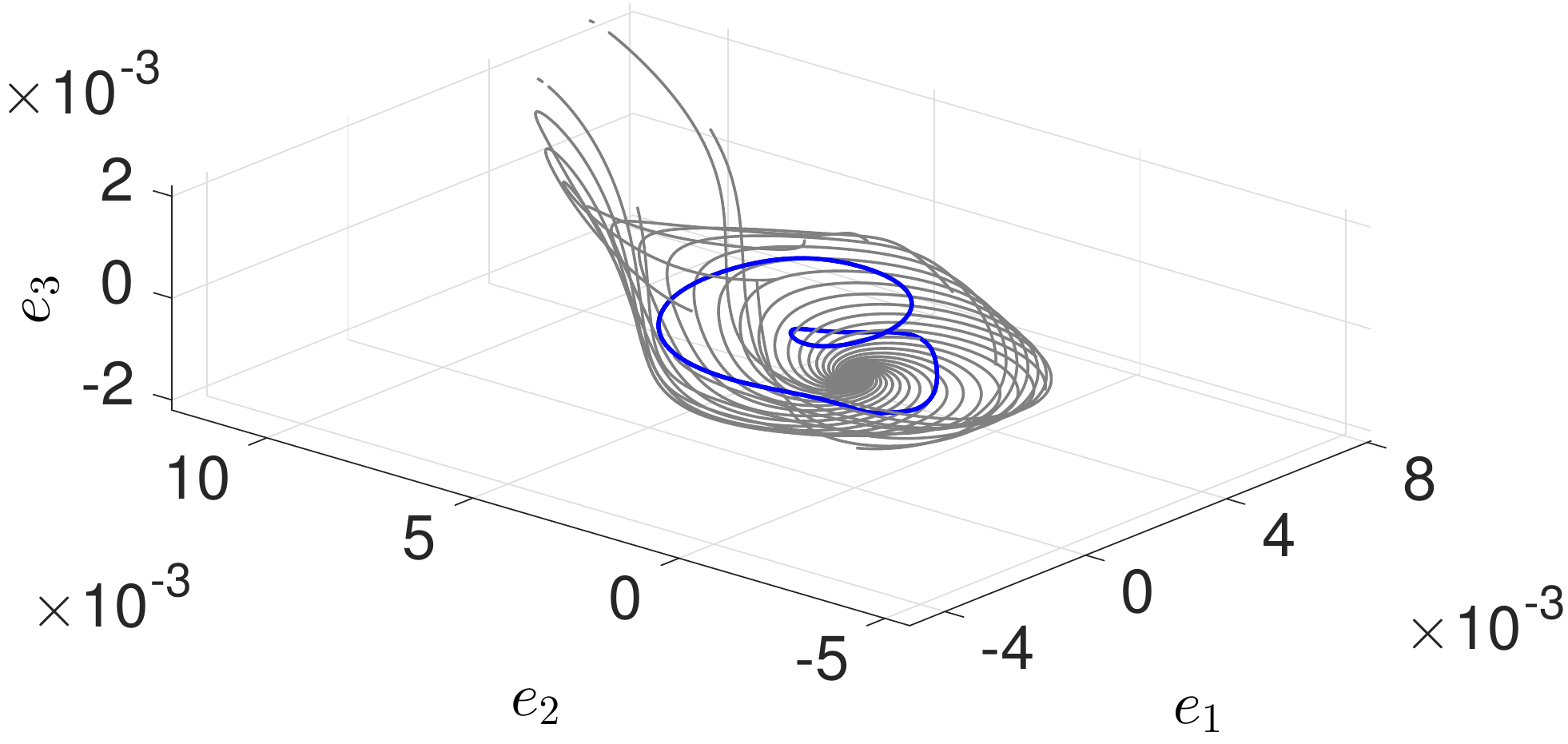}
	(b)\includegraphics[width=.42\textwidth]{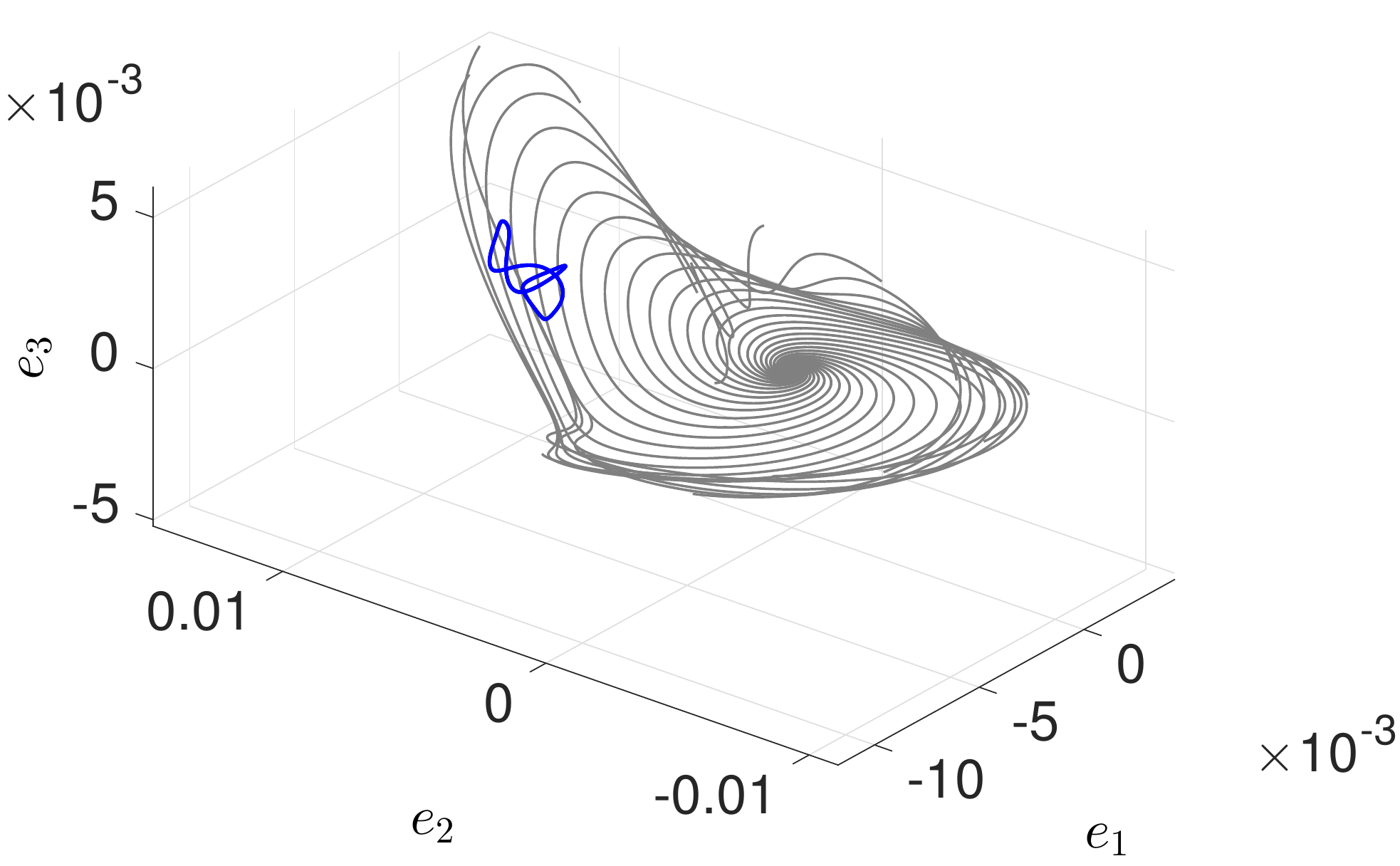}
	\caption{(Colour online)
		Low-dimensional \statesp\ visualisation of the \reqv\
        \TWRUN{1.968}{6472} (origin) and the \rpo\
		\RPORUN{11.696}{8023} (blue). A two-dimensional unstable submanifold of
		 the \reqv\ is approximated by perturbations (gray curves) around the \reqv\
         (see the text for the details). Both panels show the same objects projected to the same three-dimensional
         subspace. Two types of inner products are used for the projections:
		(a) \lowPass\ inner product \eqref{e-inprodLP},
		(b) $L^2$ inner product \eqref{e-inprodL2Data}.
	}
	\label{fig:L2vsH-1_rpo}
\end{figure}

\refFig{fig:L2vsH-1_rpo} shows a two-dimensional unstable submanifold of
\TWRUN{1.968}{6472}. Note that the unstable manifold of this \reqv\
is nine-dimensional ($d_U=9$ in \reftab{t-data}).
The visualized two-dimensional submanifold corresponds to
its most unstable subspace characterized by the largest linear stability exponent
of the \reqv. This dominant exponent is complex valued,
with a complex eigen-direction $V_1$ which defines a two-dimensional
subspace $(\Re V_1,\Im V_1)$. The three-dimensional visualizations
of \reffig{fig:L2vsH-1_rpo} are obtained by projecting each state to the subspace formed
by $(\Re V_1,\Im V_1,V_2)$ where $V_2$ is the third eigen-direction,
with a real linear stability exponent.

All computations are carried out in a slice with the \reqv\ \TWRUN{1.968}{6472}
used as the template for symmetry reduction, and placed at the origin of
the plots.
The unstable submanifold (gray curves) is approximated by forward-integrating
several small perturbations to the \reqv\ in the direction $\Re V_1$. Because
of the instability of the \reqv, the trajectories spiral away from the
origin. The spiraling nature of the trajectories is due to the complex
stability exponent. In a small neighborhood of the \reqv, the ensemble of the
trajectories approximates the two-dimensional unstable submanifold that is
tangent to the plane $(\Re V_1,\Im V_1)$. Away from the \reqv\ this
approximation fails and the trajectories diverge.

Also shown in \reffig{fig:L2vsH-1_rpo} is the \rpo\ \RPORUN{11.696}{8023}
(blue curve). The two panels show the same objects projected to the same
subspace $(\Re V_1,\Im V_1,V_2)$. The difference is that in panel ({\it a})
the \lowpass\ inner product \eqref{e-inprodLP} is used for the projection
while in panel (b) the $L^2$ inner product \eqref{e-inprodL2Data} is used.

In the \lowpass\ -projection, the \rpo\ sits near the \reqv\ and appears to be
shaped by its unstable manifold. In the $L^2$~projection, however, the \rpo\
appears to lie rather far from the \reqv, and there is no hint that their
shapes are related. We attribute this to the fact that the \lowpass\ norm
filters small scale features, assessing the distance between fluid states
based on their large-scale structures. In the $L^2$ norm, on the other hand,
even minute small-scale differences between two states contribute to the
computed distance.

We close by noting that the \lowpass\ norm was also used to detect
near-recurrences of ergodic trajectories. These near-recurrences then
served as the initial Newton iteration guesses for obtaining the \rpo s
reported in \reftab{t-data}. We find that the recurrences measured in the
\lowpass\ norm tend to converge to \rpo s more frequently than the
recurrences measured in the energy norm. A similar observation was
reported by \cite{ACHKW11}.

\section{Conclusion and perspectives}
\label{s:concl}

We investigated \rpo s embedded in transitionally turbulent
pipe flow confined to a small computational domain. These orbits were found
by Newton-type searches, using near-recurrences of the turbulent flow
as initial guesses to generate dynamically-relevant solutions.
Even in our minimal domain, made small by unphysical
symmetry restrictions, this turned out to be a daunting task,
practicable only after reduction of problem's continuous symmetry and, in
some cases, requiring also the multiple shooting Newton method.

Nonetheless, we were able to identify \NRPOsTot\ distinct \rpo s with
numerical precision of $10^{-6}$ or smaller. While this, to the best of
our knowledge, is the largest number of \po s for a three-dimensional
turbulent flow found so far, the analysis of~\refsect{s-global} shows
that only some of the \stateDsp\ visited by turbulence are populated by
the set of  \rpo s found so far.  Nevertheless, our \rpo s do occupy a
region of the \statesp\ frequently visited by turbulent trajectories, suggesting
that additional searches for \rpo s are needed to adequately represent
a larger portion of state space.
This is consistent with our expectation that in the \statesp, turbulence
is ``guided'' by the exact {\cohStr s}.

Our main result is that there is an intrinsic geometry of
turbulence, but that one has to explore the \NS\ symmetry-reduced
\statesp\ very closely in order to discern it. This geometry does not
follow from na\"{i}ve traditional statistical assumptions, as
illustrated here by the \stateDsp\ visualisations of~\refsect{s:stateSp}.
\edit{Principal component analysis (PCA)},
which we used for global projections of the dynamics
in~\refsect{s-global}, treats turbulent data as if it were a multivariate
Gaussian distribution. The true global attractor is in no sense a
Gaussian; the intrinsic geometry of turbulence revealed here
is dictated by exact time-{\cohStr s} of \NSe.

Our conclusions are not surprising to a nonlinear dynamicist experienced
in working with low-dimensional dynamical systems and their strange
attractors, yet the notion that there is an intrinsic geometry to \NS\
long-time dynamics is not as well appreciated in the turbulence
community.  This under-appreciation is historical, stemming from times
when we lacked computational tools to determine the non-trivial exact
{\cohStr s} of \NSe. In this regard, the primary contribution of this
paper is the demonstration of the computational feasibility of studying
pipe-flow turbulence as a dynamical system. For example, consider
\reffig{f-Poincare} where 147 individual turbulent runs were necessary to
obtain a very rough feeling for how turbulent trajectories are
distributed in the \stateDsp. In the dynamical systems approach, it took
$8$ carefully chosen trajectories to reveal the shape of
\RPORUN{F/6.668}{6669}'s unstable manifold.

Given the exploratory nature of this project, many of its
intermediate steps were carried out manually, yet most of these could be
automated.  The first step would be to initiate \rpo\ Newton searches by
detecting near-recurrences of turbulent flows without human supervision.
A slightly more involved step --- time-adaptive integration of symmetry-reduced
dynamics --- will probably be necessary when the azimuthal
rotation symmetry is reduced simultaneously with axial translations.
We circumvented this issue
here by restricting dynamics to the shift-and-reflect invariant
subspace, which precludes continuous rotations. This
restriction is unphysical and not present in the full problem.

Our explorations of the \stateDsp\ geometry relied on visualisations of
\rpo s and their unstable manifolds. It is already apparent from our
data in \reftab{t-data} that this strategy has limited
applicability since all but two invariant solutions we found have
unstable manifolds of dimension larger than 3. Even though partial
visualisations of the unstable manifold in~\refsect{s-local} were insightful,
there is no guarantee that this approach can extend
to larger computational domains. Ultimately, one needs to develop new methods
for systematic study of high-dimensional manifolds, a dynamical notion
of `distance' that does not depend on the particular choice of norm,
and geometric criteria for distinguishing qualitatively different
dynamics in \stateDsp.

In conclusion, we reported here our progress in dynamical study of
moderate-\Reynolds\ turbulence in the context of pipe flow. In
particular, we demonstrated that embedded within this flow are \rpo s and
that they shape dynamics in their respective neighbourhoods through their
unstable manifolds. This required various technical obstacles to be
overcome, which forced us to restrict this exploratory study to a small
symmetry-restricted computational cell. In this sense, we can say that
the dynamical approach to turbulence is still in its infancy, but the
stage is now set for study of dynamics of wall-bounded shear flow
turbulence in its full glory.

\begin{acknowledgments}
We are indebted to J.\,F.\ Gibson for many inspiring discussions.
We are grateful to  Kavli Institute for Theoretical Physics, where the
collaboration was supported in part by the National Science Foundation
under Grant No. NSF PHY11-25915, for hospitality.
A.\,P.\,W.\ was supported by the EPSRC grant EP/K03636X/1.
K.\,Y.\,S.\ was supported by the NSF Graduate
Research Fellowship Grant NSF~DGE-0707424,
P.\,C.\ was partly supported by NSF Grant DMS-0807574,
and thanks the family of G.~Robinson,~Jr.\ for support.
\end{acknowledgments}

\begin{appendices}
\section{Multi-point shooting}
\label{s:multi}

We used a multi-point shooting method (outlined in this Appendix) in
order to find some of the (relative) \po s reported in~\refsect{s:data}
(marked with subscript `M' in \reftab{t-data}). For simplicity, we
explain the concept for \po s. The approach is essentially the same for
\rpo s, once the drifts in the continuous symmetry group directions are
accounted for.

Consider a state $\ssp_0$ on a \po\ with period $T$, \ie,
\[
\ssp_0=\flow{T}{\ssp_0}
\,.
\]
The \po s are found by searching for the period $T>0$ and the state
$\ssp_0$ as the zeros of the nonlinear system of equations
$\ssp_0-\flow{T}{\ssp_0}=0$. We determine these zeros from a starting
guess by Newton--GMRES--hook iterations \citep{Visw07b}.

\begin{figure}
\centering
\includegraphics[width=.45\textwidth]{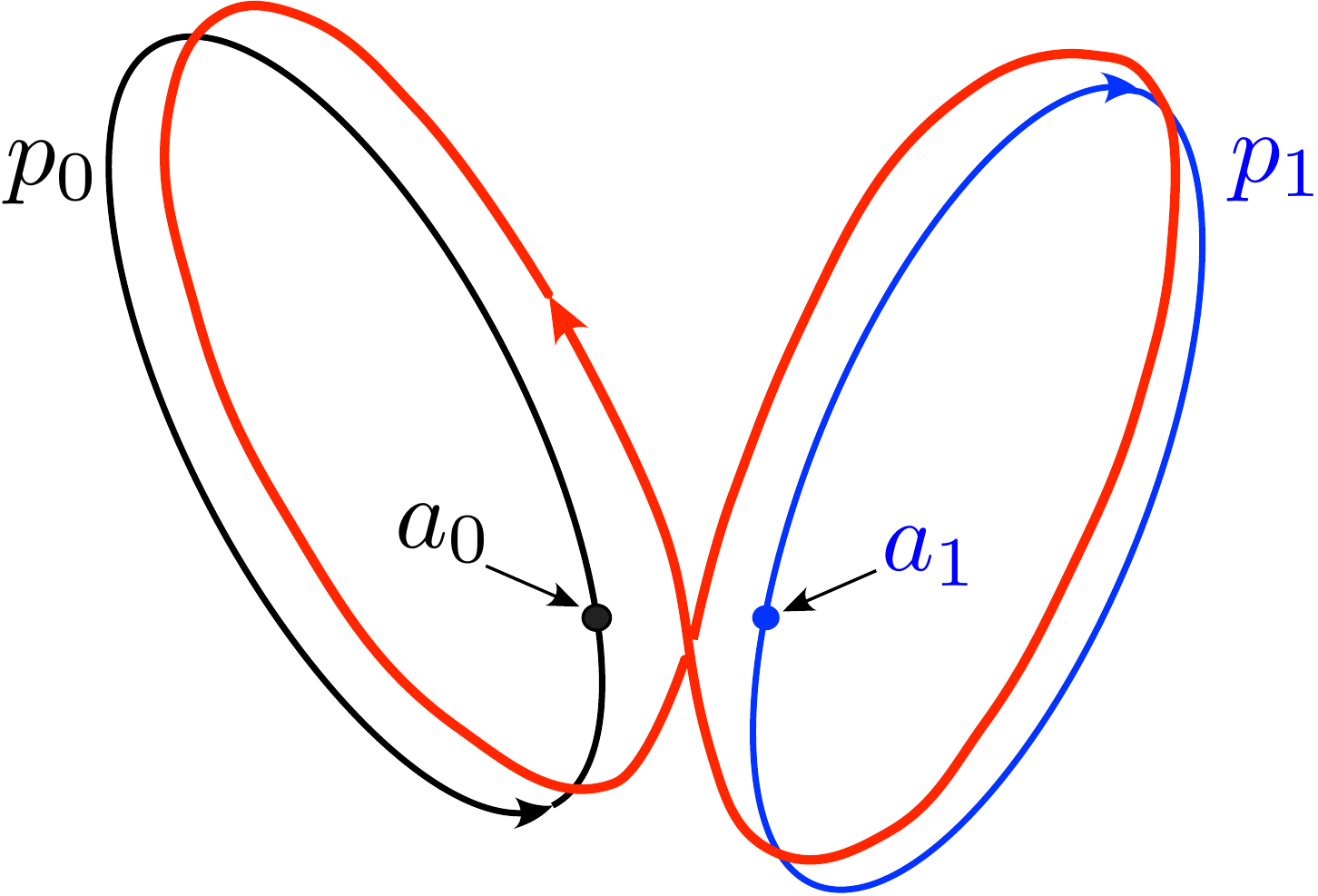}
        \caption{\label{fig:multiShoot}
        (Colour online)
An illustration of the multi-point shooting method. A long \po\ (red) is
obtained from shorter \po s $p_0$ (black) and $p_1$ (blue) with periods
$\period{0}$ and $\period{1}$, respectively. Multi-point shooting
attempts to find the shortest orbit that shadows both $p_0$ and $p_1$,
with period $\period{}$ that is approximately the sum of the periods of
the shorter orbits, $\period{}\simeq \period{0}+\period{1}$.
        }
\end{figure}

By the semi-group property of the flow map $\timeflow$,
we have
\(
\ssp_0=\flow{T}{\ssp_0}=\flow{t_2}{\flow{t_1}{\ssp_0}}
\,,
\)
for any $t_1,t_2>0$ such that $t_1+t_2=T$.
Denoting the time-$t_1$ image of $\ssp_0$ by $\ssp_1=\flow{t_1}{\ssp_0}$,
the period closes in two steps
\[
\ssp_1 =\flow{t_1}{\ssp_0}
    \,,\qquad
\ssp_0 =\flow{t_2}{a_1}
\,.
\]
The \po\ is then found through \emph{two-point shooting} by searching for
states $\ssp_0$ and $\ssp_1$ as well as the times $t_1$ and $t_2$ that
satisfy
\[
\ssp_1 -\flow{t_1}{\ssp_0}=0
    \,,\qquad
\ssp_0 -\flow{t_2}{a_1}=0\,.
\]  

The motivation for using the multi-point shooting is two-fold:
\begin{enumerate}
\item
Let $\ssp_0$ be a point on a \po\ with period $T$. In theory, we have
$\flow{T}{\ssp_0}-\ssp_0=0$. In practice, $\ssp_0$ is never known exactly
and only a numerical approximation of it is available. If the \po\ is
highly unstable, the initial discrepancy grows over time to such extent
that the state $\flow{T}{\ssp_0}$ might land far away from $\ssp_0$. By
partitioning the orbit into shorter segments, the error growth is kept
under control.

\item
Multi-point shooting offers a systematic way to create long orbits by
``glueing'' the already known shorter orbits\rfp{pchaot}. Let $p_0$ and
$p_1$ denote two short \po s with periods $T_0$ and $T_1$, respectively
(see \reffig{fig:multiShoot}). The initial guesses for $\ssp_0$ and $\ssp_1$
are chosen to belong to the shorter orbits, \ie, $\ssp_0\in p_0$ and
$\ssp_1\in p_1$. The initial guesses for the flight times $t_1$ and $t_2$
are chosen to coincide with the period of the short orbits, \ie,
$t_1=T_0$ and $t_2=T_1$. If the Newton--GMRES--hook steps converge, the
resulting orbit shadows the short \po s and has a period $T\simeq
T_0+T_1$.
\end{enumerate}

\bibliographystyle{jfm}
\bibliography{../../bibtex/pipes}

\begin{thebibliography}{63}
\expandafter\ifx\csname natexlab\endcsname\relax\def\natexlab#1{#1}\fi

\bibitem[Aubry {\em et~al.\/}(1988)Aubry, Holmes, Lumley \& Stone]{Aubry88}
{\sc Aubry, N., Holmes, P., Lumley, J.~L. \& Stone, E.} 1988 The dynamics of
  coherent structures in the wall region of a turbulent boundary layer. {\em J.
  Fluid Mech.\/} {\bf 192}, 115.

\bibitem[Auerbach {\em et~al.\/}(1987)Auerbach, Cvitanovi{\'c}, Eckmann,
  Gunaratne \& Procaccia]{pchaot}
{\sc Auerbach, D., Cvitanovi{\'c}, P., Eckmann, J.-P., Gunaratne, G. \&
  Procaccia, I.} 1987 Exploring chaotic motion through periodic orbits. {\em
  Phys. Rev. Lett.\/} {\bf 58}, 23.

\bibitem[Avila {\em et~al.\/}(2011)Avila, Moxey, de~Lozar, Avila, Barkley \&
  Hof]{AMdABH11}
{\sc Avila, K., Moxey, D., de~Lozar, A., Avila, M., Barkley, D. \& Hof, B.}
  2011 The onset of turbulence in pipe flow. {\em Science\/} {\bf 333},
  192--196.

\bibitem[Avila {\em et~al.\/}(2013)Avila, Mellibovsky, Roland \&
  Hof]{AvMeRoHo13}
{\sc Avila, M., Mellibovsky, F., Roland, N. \& Hof, B.} 2013
  Streamwise-localized solutions at the onset of turbulence in pipe flow. {\em
  Phys. Rev. Lett.\/} {\bf 110}, 224502.

\bibitem[Avila {\em et~al.\/}(2010)Avila, Willis \& Hof]{AVWIHO10}
{\sc Avila, M., Willis, A.~P. \& Hof, B.} 2010 On the transient nature of
  localized pipe flow turbulence. {\em J. Fluid Mech.\/} {\bf 646}, 127--136.

\bibitem[Benedicks \& Carleson(1991)]{BenCar91}
{\sc Benedicks, M. \& Carleson, L.} 1991 The dynamics of the {H{\'e}non} map.
  {\em Ann. Math.\/} {\bf 133}, 73.

\bibitem[Berkooz {\em et~al.\/}(1993)Berkooz, Holmes \& Lumley]{Berkooz93}
{\sc Berkooz, G., Holmes, P. \& Lumley, J.~L.} 1993 The proper orthogonal
  decomposition in the analysis of turbulent flows. {\em Ann. Rev. Fluid
  Mech.\/} {\bf 25}, 539--575.

\bibitem[Budanur(2015)]{BudanurThesis}
{\sc Budanur, N.~B.} 2015 Exact coherent structures in spatiotemporal chaos:
  From qualitative description to quantitative predictions. PhD thesis, School
  of Physics, Georgia Inst. of Technology, Atlanta.

\bibitem[Budanur \& Cvitanovi\'c(2015)]{BudCvi15}
{\sc Budanur, N.~B. \& Cvitanovi\'c, P.} 2015 Unstable manifolds of relative
  periodic orbits in the symmetry-reduced state space of the
  {Kuramoto-Sivashinsky} system. {\em J. Stat. Phys.\/} {\bf 167}, 636--655.

\bibitem[Budanur {\em et~al.\/}(2015)Budanur, Cvitanovi\'c, Davidchack \&
  Siminos]{BudCvi14}
{\sc Budanur, N.~B., Cvitanovi\'c, P., Davidchack, R.~L. \& Siminos, E.} 2015
  Reduction of the {SO(2)} symmetry for spatially extended dynamical systems.
  {\em Phys. Rev. Lett.\/} {\bf 114}, 084102.

\bibitem[Budanur \& Hof(2017)]{BudHof17}
{\sc Budanur, N.~B. \& Hof, B.} 2017 Heteroclinic path to spatially localized
  chaos in pipe flow.

\bibitem[de~Carvalho \& Hall(2002)]{dCH02}
{\sc de~Carvalho, A. \& Hall, T.} 2002 How to prune a horseshoe. {\em
  Nonlinearity\/} {\bf 15}, R19–R68.

\bibitem[Chandler \& Kerswell(2013)]{ChaKer12}
{\sc Chandler, G.~J. \& Kerswell, R.~R.} 2013 Invariant recurrent solutions
  embedded in a turbulent two-dimensional {Kolmogorov} flow. {\em J. Fluid
  M.\/} {\bf 722}, 554--595, \arXiv{1207.4682}.

\bibitem[Cvitanovi{\'c}(2017)]{statespDummies}
{\sc Cvitanovi{\'c}, P.} 2017 Life in extreme dimensions. In {\em {Chaos:
  Classical and Quantum}\/} (ed. P.~Cvitanovi{\'c}, R.~Artuso, R.~Mainieri,
  G.~Tanner \& G.~Vattay). Copenhagen: Niels Bohr Inst.

\bibitem[Cvitanovi{\'c} {\em et~al.\/}(2017)Cvitanovi{\'c}, Artuso, Mainieri,
  Tanner \& Vattay]{DasBuch}
{\sc Cvitanovi{\'c}, P., Artuso, R., Mainieri, R., Tanner, G. \& Vattay, G.}
  2017 {\em Chaos: Classical and Quantum\/}. Copenhagen: Niels Bohr Inst.

\bibitem[Cvitanovi\'c {\em et~al.\/}(2012)Cvitanovi\'c, Borrero-Echeverry,
  Carroll, Robbins \& Siminos]{atlas12}
{\sc Cvitanovi\'c, P., Borrero-Echeverry, D., Carroll, K., Robbins, B. \&
  Siminos, E.} 2012 Cartography of high-dimensional flows: {A} visual guide to
  sections and slices. {\em Chaos\/} {\bf 22}, 047506.

\bibitem[Cvitanovi{\'c} \& Gibson(2010)]{CviGib10}
{\sc Cvitanovi{\'c}, P. \& Gibson, J.~F.} 2010 Geometry of turbulence in
  wall-bounded shear flows: {Periodic} orbits. {\em Phys. Scr. T\/} {\bf 142},
  014007.

\bibitem[Cvitanovi\'{c} {\em et~al.\/}(1988)Cvitanovi\'{c}, Gunaratne \&
  Procaccia]{pre88top}
{\sc Cvitanovi\'{c}, P., Gunaratne, G.~H. \& Procaccia, I.} 1988 Topological
  and metric properties of {H\'enon}-type strange attractors. {\em Phys. Rev.
  A\/} {\bf 38}, 1503.

\bibitem[Dennis \& Sogaro(2014)]{DeSo14}
{\sc Dennis, D. J.~C. \& Sogaro, F.~M.} 2014 Distinct organizational states of
  fully developed turbulent pipe flow. {\em Phys. Rev. Lett.\/} {\bf 113},
  234501.

\bibitem[Duguet {\em et~al.\/}(2008)Duguet, Willis \& Kerswell]{duguet07}
{\sc Duguet, Y., Willis, A.~P. \& Kerswell, R.~R.} 2008 Transition in pipe
  flow: the saddle structure on the boundary of turbulence. {\em J. Fluid
  Mech.\/} {\bf 613}, 255--274.

\bibitem[Eggels {\em et~al.\/}(1994)Eggels, Unger, Weiss, Westerweel, Adrian,
  Freidrich \& Nieuwstadt]{Eggels94}
{\sc Eggels, J. G.~M., Unger, F., Weiss, M.~H., Westerweel, J., Adrian, R.~J.,
  Freidrich, R. \& Nieuwstadt, F. T.~M.} 1994 Fully developed turbulent pipe
  flow: a comparison between direct numberical simulation and experiment. {\em
  J. Fluid Mech.\/} {\bf 268}, 175--209.

\bibitem[Faisst \& Eckhardt(2003)]{FE03}
{\sc Faisst, H. \& Eckhardt, B.} 2003 Traveling waves in pipe flow. {\em Phys.
  Rev. Lett.\/} {\bf 91}, 224502.

\bibitem[Farazmand(2016)]{Faraz15}
{\sc Farazmand, M.} 2016 An adjoint-based approach for finding invariant
  solutions of {Navier-Stokes} equations. {\em J. Fluid Mech.\/} {\bf 795},
  278--312.

\bibitem[Gibson(2017)]{channelflow}
{\sc Gibson, J.~F.} 2017 {Channelflow}: {A} spectral {Navier-Stokes} simulator
  in {C}++. {\em Tech. Rep.\/}. U. New Hampshire, {\tt {Channelflow.org}}.

\bibitem[Gibson {\em et~al.\/}(2008)Gibson, Halcrow \& Cvitanovi{\'c}]{GHCW07}
{\sc Gibson, J.~F., Halcrow, J. \& Cvitanovi{\'c}, P.} 2008 Visualizing the
  geometry of state space in plane {Couette} flow. {\em J. Fluid Mech.\/} {\bf
  611}, 107--130.

\bibitem[Gibson {\em et~al.\/}(2009)Gibson, Halcrow \& Cvitanovi{\'c}]{HGC08}
{\sc Gibson, J.~F., Halcrow, J. \& Cvitanovi{\'c}, P.} 2009 Equilibrium and
  traveling-wave solutions of plane {Couette} flow. {\em J. Fluid Mech.\/} {\bf
  638}, 243--266.

\bibitem[Hagen(1839)]{Hagen1839}
{\sc Hagen, G.} 1839 {\"U}ber die bewegung des wassers in engen cylindrischen
  r{\"o}hren. {\em Ann. Phys.\/} {\bf 122}, 423--442.

\bibitem[Halcrow {\em et~al.\/}(2009)Halcrow, Gibson, Cvitanovi{\'c} \&
  Viswanath]{GHCV08}
{\sc Halcrow, J., Gibson, J.~F., Cvitanovi{\'c}, P. \& Viswanath, D.} 2009
  Heteroclinic connections in plane {Couette} flow. {\em J. Fluid Mech.\/} {\bf
  621}, 365--376.

\bibitem[Hamilton {\em et~al.\/}(1995)Hamilton, Kim \& Waleffe]{HaKiWa95}
{\sc Hamilton, J.~M., Kim, J. \& Waleffe, F.} 1995 Regeneration mechanisms of
  near-wall turbulence structures. {\em J. Fluid Mech.\/} {\bf 287}, 317--348.

\bibitem[Hof {\em et~al.\/}(2004)Hof, van Doorne, Westerweel, Nieuwstadt,
  Faisst, Eckhardt, Wedin, Kerswell \& Waleffe]{science04}
{\sc Hof, B., van Doorne, C. W.~H., Westerweel, J., Nieuwstadt, F. T.~M.,
  Faisst, H., Eckhardt, B., Wedin, H., Kerswell, R.~R. \& Waleffe, F.} 2004
  Experimental observation of nonlinear traveling waves in turbulent pipe flow.
  {\em Science\/} {\bf 305}, 1594--1598.

\bibitem[Hof {\em et~al.\/}(2006)Hof, Westerweel, Schneider \&
  Eckhardt]{hof2006flt}
{\sc Hof, B., Westerweel, J., Schneider, T.~M. \& Eckhardt, B.} 2006 Finite
  lifetime of turbulence in shear flows. {\em Nature\/} {\bf 443}~(7107),
  59--62.

\bibitem[Holmes {\em et~al.\/}(1996)Holmes, Lumley \& Berkooz]{Holmes96}
{\sc Holmes, P., Lumley, J.~L. \& Berkooz, G.} 1996 {\em {Turbulence, Coherent
  Structures, Dynamical Systems and Symmetry}\/}. Cambridge: Cambridge Univ.
  Press.

\bibitem[Hopcroft \& Kannan(2014)]{HopKan14}
{\sc Hopcroft, J. \& Kannan, R.} 2014 {Foundations of Data Science}. In
  preparation.

\bibitem[Hopf(1948)]{Hopf48}
{\sc Hopf, E.} 1948 A mathematical example displaying features of turbulence.
  {\em Commun. Pure Appl. Math.\/} {\bf 1}, 303--322.

\bibitem[Jim{\'e}nez \& Moin(1991)]{JM91}
{\sc Jim{\'e}nez, J. \& Moin, P.} 1991 The minimal flow unit in near-wall
  turbulence. {\em J. Fluid Mech.\/} {\bf 225}, 213--240.

\bibitem[Kawahara \& Kida(2001)]{KawKida01}
{\sc Kawahara, G. \& Kida, S.} 2001 Periodic motion embedded in plane {Couette}
  turbulence: {Regeneration} cycle and burst. {\em J. Fluid Mech.\/} {\bf 449},
  291--300.

\bibitem[Kerswell \& Tutty(2007)]{KeTu06}
{\sc Kerswell, R.~R. \& Tutty, O.} 2007 Recurrence of travelling waves in
  transitional pipe flow. {\em J. Fluid Mech.\/} {\bf 584}, 69--102.

\bibitem[Kreilos \& Eckhardt(2012)]{KreEck12}
{\sc Kreilos, T. \& Eckhardt, B.} 2012 Periodic orbits near onset of chaos in
  plane {Couette} flow. {\em Chaos\/} {\bf 22}, 047505.

\bibitem[Kreilos {\em et~al.\/}(2014)Kreilos, Zammert \& Eckhardt]{KrZaEc14}
{\sc Kreilos, T., Zammert, S. \& Eckhardt, B.} 2014 Comoving frames and
  symmetry-related motions in parallel shear flows. {\em J. Fluid Mech.\/} {\bf
  751}, 685--697.

\bibitem[Lax(2002)]{Lax02}
{\sc Lax, P.~D.} 2002 {\em {Functional Analysis}\/}. New York: Wiley.

\bibitem[Lin {\em et~al.\/}(2011)Lin, Thiffeault \& Doering]{Lin2011}
{\sc Lin, Z., Thiffeault, J.-L. \& Doering, C.~R.} 2011 Optimal stirring
  strategies for passive scalar mixing. {\em J. Fluid Mech.\/} {\bf 675},
  465--476.

\bibitem[Mathew {\em et~al.\/}(2007)Mathew, Mezi{\'C}, Grivopoulos, Vaidya \&
  Petzold]{Mathew07}
{\sc Mathew, G., Mezi{\'C}, I., Grivopoulos, S., Vaidya, U. \& Petzold, L.}
  2007 Optimal control of mixing in {Stokes} fluid flows. {\em J. Fluid
  Mech.\/} {\bf 580}, 261--281.

\bibitem[Mellibovsky \& Eckhardt(2011)]{mellibovsky11}
{\sc Mellibovsky, F. \& Eckhardt, B.} 2011 {Takens--Bogdanov} bifurcation of
  travelling-wave solutions in pipe flow. {\em J. Fluid Mech.\/} {\bf 670},
  96--129.

\bibitem[Mellibovsky \& Eckhardt(2012)]{mellibovsky12}
{\sc Mellibovsky, F. \& Eckhardt, B.} 2012 From travelling waves to mild chaos:
  {A} supercritical bifurcation cascade in pipe flow. {\em J. Fluid Mech.\/}
  {\bf 709}, 149--190.

\bibitem[Meseguer \& Trefethen(2003)]{MesTre03}
{\sc Meseguer, A. \& Trefethen, L.~N.} 2003 Linearized pipe flow to {Reynolds}
  number $10^7$. {\em J. Comput. Phys.\/} {\bf 186}, 178--197.

\bibitem[Poiseuille(1840)]{Poiseuille1844}
{\sc Poiseuille, J.~L.} 1840 Recherches exp{\'e}rimentales sur le mouvement des
  liquides dans les tubes de tr{\`e}s-petits diam{\`e}tres. {\em C. R. Acad.
  Sci.\/} {\bf 11}, 961.

\bibitem[Pringle {\em et~al.\/}(2009)Pringle, Duguet \& Kerswell]{Pringle09}
{\sc Pringle, C. C.~T., Duguet, Y. \& Kerswell, R.~R.} 2009 Highly symmetric
  travelling waves in pipe flow. {\em Phil. Trans. Royal Soc. A\/} {\bf 367},
  457--472.

\bibitem[Pringle \& Kerswell(2007)]{Pringle07}
{\sc Pringle, C. C.~T. \& Kerswell, R.~R.} 2007 Asymmetric, helical, and
  mirror-symmetric traveling waves in pipe flow. {\em Phys. Rev. Lett.\/} {\bf
  99}, 074502.

\bibitem[Reynolds(1894)]{R1894}
{\sc Reynolds, O.} 1894 On the dynamical theory of incompressible viscous flows
  and the determination of the criterion. {\em Proc. Roy. Soc. Lond. Ser. A\/}
  {\bf 186}, 123--161.

\bibitem[Rowley {\em et~al.\/}(2003)Rowley, Kevrekidis, Marsden \&
  Lust]{rowley_reduction_2003}
{\sc Rowley, C.~W., Kevrekidis, I.~G., Marsden, J.~E. \& Lust, K.} 2003
  Reduction and reconstruction for self-similar dynamical systems. {\em
  Nonlinearity\/} {\bf 16}, 1257--1275.

\bibitem[Rowley \& Marsden(2000)]{rowley_reconstruction_2000}
{\sc Rowley, C.~W. \& Marsden, J.~E.} 2000 Reconstruction equations and the
  {Karhunen-Lo{\'e}ve} expansion for systems with symmetry. {\em Physica D\/}
  {\bf 142}, 1--19.

\bibitem[Schmiegel(1999)]{Schmi99}
{\sc Schmiegel, A.} 1999 Transition to turbulence in linearly stable shear
  flows. PhD thesis, Philipps-Universit{\"a}t Marburg, available on {\tt
  archiv.ub.uni-marburg.de/diss/z2000/0062}.

\bibitem[Schmiegel \& Eckhardt(1997)]{SE97}
{\sc Schmiegel, A. \& Eckhardt, B.} 1997 Fractal stability border in plane
  {Couette} flow. {\em Phys. Rev. Lett.\/} {\bf 79}, 5250.

\bibitem[Schneider {\em et~al.\/}(2007{\natexlab{{\em a\/}}})Schneider,
  Eckhardt \& Vollmer]{SchEckVoll07}
{\sc Schneider, T.~M., Eckhardt, B. \& Vollmer, J.} 2007{\natexlab{{\em a\/}}}
  Statistical analysis of coherent structures in transitional pipe flow. {\em
  Phys. Rev. E\/} {\bf 75}, 066313.

\bibitem[Schneider {\em et~al.\/}(2007{\natexlab{{\em b\/}}})Schneider,
  Eckhardt \& Yorke]{SchEckYor07}
{\sc Schneider, T.~M., Eckhardt, B. \& Yorke, J.} 2007{\natexlab{{\em b\/}}}
  Turbulence, transition, and the edge of chaos in pipe flow. {\em Phys. Rev.
  Lett.\/} {\bf 99}, 034502.

\bibitem[Short \& Willis(2017)]{ShWi16}
{\sc Short, K.~Y. \& Willis, A.~P.} 2017 Bifurcation structure of relative
  periodic orbits in pipe flow. In preparation.

\bibitem[Spruill(2007)]{Spruill07}
{\sc Spruill, M.~C.} 2007 Asymptotic distribution of coordinates on high
  dimensional spheres. {\em Elect. Commun. in Probab.\/} {\bf 12}, 234--247.

\bibitem[Viswanath(2007)]{Visw07b}
{\sc Viswanath, D.} 2007 Recurrent motions within plane {Couette} turbulence.
  {\em J. Fluid Mech.\/} {\bf 580}, 339--358.

\bibitem[Wedin \& Kerswell(2004)]{WK04}
{\sc Wedin, H. \& Kerswell, R.~R.} 2004 Exact coherent structures in pipe flow:
  {Traveling} wave solutions. {\em J. Fluid Mech.\/} {\bf 508}, 333--371.

\bibitem[Wegman \& Solka(2002)]{WegSol02}
{\sc Wegman, E.~J. \& Solka, J.~L.} 2002 On some mathematics for visualizing
  high dimensional data. {\em Sankhy{\=a}: Indian J. Statistics, Ser. A\/} pp.
  429--452.

\bibitem[Willis {\em et~al.\/}(2013)Willis, Cvitanovi{\'c} \& Avila]{ACHKW11}
{\sc Willis, A.~P., Cvitanovi{\'c}, P. \& Avila, M.} 2013 Revealing the state
  space of turbulent pipe flow by symmetry reduction. {\em J. Fluid Mech.\/}
  {\bf 721}, 514--540.

\bibitem[Willis \& Kerswell(2008)]{WillKer08}
{\sc Willis, A.~P. \& Kerswell, R.~R.} 2008 Coherent structures in localised
  and global pipe turbulence. {\em Phys. Rev. Lett.\/} {\bf 100}, 124501.

\bibitem[Willis {\em et~al.\/}(2016)Willis, Short \& Cvitanovi{\'c}]{WiShCv15}
{\sc Willis, A.~P., Short, K.~Y. \& Cvitanovi{\'c}, P.} 2016 Symmetry reduction
  in high dimensions, illustrated in a turbulent pipe. {\em Phys. Rev. E\/}
  {\bf 93}, 022204.

\end{thebibliography}

\end{appendices}

\end{document}